\documentclass[9pt]{elife}

\usepackage{lipsum} 
\usepackage[version=4]{mhchem}
\usepackage{siunitx}
\usepackage{multirow}

\newcommand{\beginsupplement}{%
        \setcounter{table}{0}
        \renewcommand{\thetable}{S\arabic{table}}%
        \setcounter{figure}{0}
        \renewcommand{\thefigure}{S\arabic{figure}}%
     }

\DeclareSIUnit\Molar{M}

\title{Individual crop loads provide local control for collective food intake in ant colonies}

\author[1\authfn{1}]{Efrat Greenwald}
\author[1\authfn{1}]{Lior Baltiansky}
\author[1*]{Ofer Feinerman}

\affil[1]{Department of Physics of Complex Systems, Weizmann Institute of Science, Rehovot, Israel.}

\corr{ofer.feinerman@weizmann.ac.il}{}

\contrib[\authfn{1}]{These authors contributed equally to this work}

\begin{document}

\maketitle

\begin{abstract}
    Nutritional regulation by ants emerges from a distributed process: food is collected by a small fraction of workers, stored within the crops of individuals, and spreads via local ant-to-ant interactions. The precise individual-level underpinnings of this collective regulation have remained unclear mainly due to difficulties in measuring food within ants’ crops. Here we image fluorescent liquid food in individually tagged \textit{Camponotus sanctus} ants, and track the real-time food flow from foragers to their gradually satiating colonies. We show how the feedback between colony satiation level and food inflow is mediated  by individual crop loads;  specifically, the crop loads of recipient ants control food flow rates, while those of foragers regulate the frequency of foraging-trips. Interestingly, these effects do not rise from pure physical limitations of crop capacity. Our findings suggest that the emergence of food intake regulation does not require individual foragers to assess the global state of the colony.
\end{abstract}

\section*{Introduction}

Eusocial insects stand out in their ability to achieve collective regulation with no central control. Nutritional management in bees and ants is a compelling example. On the one hand, the colony as a whole displays high levels of collective regulation on the amount of food collected \cite{howard1980effect,sorensen1985control,cassill1999regulation}, on its nutritional composition \cite{dussutour2009communal,cook2010colony,bazazi2016responses}, and on its dissemination within the colony \cite{anderson1999task,sendova2010emergency,greenwald2015ant}. On the other hand, this regulation is achieved by individuals that react to their local environment. Food dissemination often relies on local \textit{trophallactic} interactions in which liquid food, not fully digested, is regurgitated from the crop of one individual and passed mouth-to-mouth to another (\cite{holldobler1990ants} chapter 7, page 291 and SI of \cite{greenwald2015ant}). Such distributed processes are characterized by intricate interaction networks that include significant random aspects \cite{fewell2003social,pinter2011effect,mersch2013tracking,sendova2010emergency} which may hinder global coordination. How colonies manage to achieve tight nutritional regulation despite the difficulties that are inherent to a distributed 
process is not fully understood.

An essential component in the  nutritional regulation of any living system is the adjustment of incoming food rates to the current level of satiation \cite{parks2012theory,simpson1993multi,josens2000foraging}. To experimentally approach the principles behind such adjustments it is useful to observe the global process in which food accumulates in the system. Indeed, the dynamics of food accumulation in ant colonies have been a subject of interest for many years \cite{wilson1957quantitative,markin1970food,howard1980effect,buffin2009feeding,buffin2012collective,sendova2010emergency,greenwald2015ant}.
These studies show that when introduced to a new food source, the levels of food stored within the colony display logistic dynamics. The logistic growth in the amount of accumulated food supports the notion that total food inflow is regulated by the amount of food already stored within the colony. The local origins of this global regulation are still not fully understood.

To understand how global food flow regulation emerges from single ant behaviors one should consider the forager ants. These ants, which typically constitute only a small fraction of all workers, are the ones responsible for bringing food into the nest \cite{oster1978caste,traniello1977recruitment}. Therefore, any change in the global inflow of food to the colony must be manifested in the rate at which foragers collect and deliver food. Accordingly, colonies can regulate the inflow of food by modulating foraging effort: for example, by varying the number of active foragers through recruitment \cite{gordon2002regulation}. Indeed, many studies on the regulation of foraging have focused on recruitment behavior and have shown that it correlates with the colony's nutritional state \cite{traniello1977recruitment,seeley1989social,tenczar2014automated,cassill2003rules}. In this work, we explore a less studied aspect of food flow regulation, namely, changes in the behavior of already active foragers. Active foragers engage in repeated trips between the food source and the nest \cite{traniello1977recruitment,tenczar2014automated},
where they use trophallaxis to deliver their food load to multiple recipients \cite{seeley1989social,gregson2003partial,huang2003multiple,traniello1977recruitment}. The rate at which a forager leaves the nest for her next trip as well as the amount of food that she manages to unload per trip provide potential regulators of the collective foraging effort. These regulators may be tied to the colony's nutritional state through the experience of returning foragers when they unload in the nest. In this vein, it was shown that honeybee foragers experience longer waiting times between subsequent unloading interactions if the colony is satiated \cite{seeley1989social}, and it was suggested that they use this information to adjust their recruitment behavior.

Most previous studies of individual forager behavior did not make direct connections between single ant rules and the global dynamics of food accumulation \cite{seeley1989social,huang2003multiple,gregson2003partial}. Nonetheless, the observations and interpretations they present  are consistent with a simple intuition for the origins of the observed logistic dynamics in the accumulation of food: Initially, when a scout from a hungry colony encounters food she commences a recruitment process in which the number of active foragers increases \cite{greene2007interaction}. This positive feedback is followed by a delayed negative feedback that results from  the increased difficulty that foragers have in locating available recipients as  the colony  satiates \cite{seeley1989social,Seeley1994-xw, buffin2009feeding,sendova2010emergency}. A simple prediction follows: if foragers unload their entire crop contents before leaving for their next foraging trip \cite{gregson2003partial,traniello1977recruitment}, the frequency at which a forager exits the nest should gradually decrease as the colony satiates \cite{buffin2009feeding}.

Although the above intuition may seem complete, it has only little empirical support.
Until recently,  microscopic measurements of real-time individual crop loads and food-flows in single interactions were unavailable. As a result, existing explanations for different aspects of the foraging process rely (either explicitly or implicitly) on various assumptions. The foragers were assumed to unload their entire crop contents before leaving the nest \cite{traniello1977recruitment,gregson2003partial,buffin2009feeding} and use local experience to assess the colony's nutritional state
\cite{seeley1989social,Seeley1994-xw,huang2003multiple}. The recipients were assumed to be either empty or full  \cite{sendova2010emergency,seeley1989social, Seeley1994-xw} and  fill upon a single interaction with a forager \cite{sendova2010emergency,seeley1989social, Seeley1994-xw}. As for the pattern of interactions between foragers and their recipients, it was assumed that in the nest a forager has a constant probability per unit time to interact with potential recipients \cite{sendova2010emergency,seeley1989social, Seeley1994-xw}, there is a formation of queues of returning foragers and available receivers \cite{seeley1989social}, and that interaction patterns are random \cite{Seeley1994-xw,buffin2009feeding,sendova2010emergency}.

Relying on individual-level assumptions may be deceiving since  multiple sets of microscopic rules can lead to similar macroscopic outcomes. For example, the slowing down of foragers' unloading rates may stem from reduced rates of trophallactic interactions but can also be the result of smaller amounts of food transferred per interaction. Both will affect the global outcome similarly. To uniquely identify the micro-scale mechanisms of food inflow regulation and examine the assumptions outlined above, we tracked fluorescently-labeled food in crops of individually tagged ants \cite{greenwald2015ant}. This technology allowed for a non-intrusive study of the dynamics of food accumulation in ant colonies 
with a spatial resolution of single-ant crop loads and a temporal resolution sufficient to capture single trophallactic events. We thus present the missing experimental data on the crop contents of encountered ants, the amount of food transferred per interaction, the dynamics of forager unloading at different satiety states of the colony, and the amount of food in the foragers' crops when they exit the nest.

In the following sections we use these highly resolved measurements to  quantitatively link the microscopic and macroscopic scales of food accumulation dynamics in ant colonies. We demonstrate how the global dynamics and the regulation of individual foraging effort rely on individual crop loads. Specifically, we delineate how individual crop loads affect a forager's unloading rate as well as her decision to exit the nest for the next food collection trip. Our findings suggest a distributed regulation mechanism which does not require  individual foragers to assess global, colony-scale variables.

\section*{Results}

\subsection*{Dynamics of Food Accumulation}
\begin{figure}
    \centering
    \includegraphics[scale=0.42]{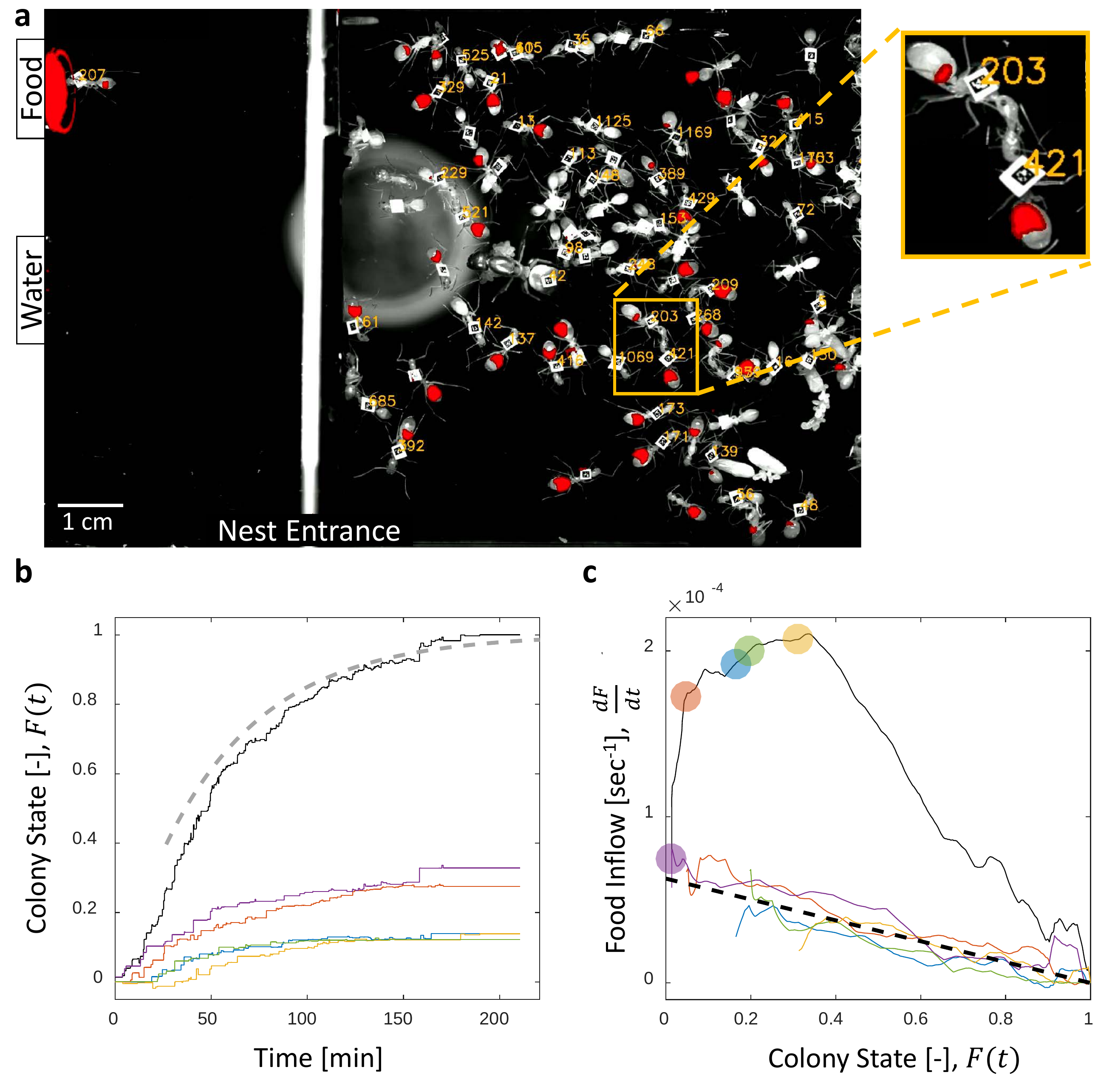}

    \caption{\textbf{Dynamics of Food Accumulation in a Starved Colony.} \textbf{(a)} A single frame from a video of a colony in the course of food accumulation. Ant identity is presented as a unique number next to her tag, and the fluorescent food is presented in red. The right side of the image is an IR-covered nest, and the left side is a neighboring open yard that includes a food source and a water source. In this frame a forager can be seen feeding from the food source (ant 207), and a trophallactic event between ants 203 and 421 is magnified. \textbf{(b)} Global food accumulation (normalized fluorescence), $F(t)$, is plotted in black. The accumulated food brought by each forager, $f_i(t)$ is plotted in a unique color. The dashed line is the predicted colony state according to equation \ref{eq. dFdt} for $m=0.6\times10^{-4}$, as estimated from Fig. \ref{fig:global}c. \textbf{(c)} The time-averaged global inflow, $\frac{dF}{dt}$, as a function of $F(t)$, is plotted in the black solid line. Time-averaged flows through individual foragers, $\frac{df_i}{dt}$, are plotted in unique colors (same as in panel b). Flows were calculated by differentiating the colony state and the contributions of each forager (the curves from Fig. \ref{fig:global}b) with respect to time (see Methods, Data Analysis). Colored circles on the global plot depict each forager's first return from the food source. The black dashed line represents equation \ref{eq. dfdt_i}, where $m$ was calculated as follows: the flow through each forager was fit with an equation of the form $\frac{df_i}{dt}=m_i\cdot (1-F(t))$, and $m$ was taken to be the average of all $m_i$. Results from all three experimental colonies can be found in Fig. \ref{fig:globalSI} and Fig. \ref{fig:averageforager}. Source files for panels b and c are available in Figure 1—source data 1 and 2.}
    \label{fig:global}
\end{figure}

Food accumulation dynamics were studied by introducing starved colonies of \textit{Camponotus sanctus} ants to fluorescently-labeled food. Food was supplied \textit{ad-libitum} to isolate the effects of colony satiation from the effect of resource availability on the inflow of food.
As the colony replenished, we followed the traffic and storage of the  food within the crops of individual ants using real-time fluorescent imaging (Fig. \ref{fig:global}a, Vid. 1 and Vid. 2, for details see Methods).

We found that the total amount of food in the colony gradually accumulated until saturation (Fig. \ref{fig:global}b, \cite{buffin2009feeding}). The level at which food saturated was defined post-hoc as the colony's intake volume target. We define the ``colony state" at time $t$, denoted $F(t)$, as the total amount of food in the colony at time $t$ divided by the colony's intake target. The ``colony state" is thus a normalized measure of the colony's satiety level, starting from $F=0$ when the colony is starved and gradually approaching $F=1$ as the colony approaches its target (Fig. \ref{fig:global}b, black line).

To enable tracing of the food flow process on the single-ant level, all ants were individually tagged (Fig. \ref{fig:global}a, and see Methods, Experimental setup). As could be expected \cite{gordon1989dynamics,tenczar2014automated}, this labeling showed that a few consistent foragers were accountable for the transfer of food from the source to the ants in the nest (Fig. \ref{fig:global}b). This allowed us to study the dynamics of food accumulation by expressing colony state as the sum of the contributions of individual foragers:

\begin{equation}
    F(t)=\sum_{i=1}^{N} f_i(t)
    \label{eq. Sum_fi}
\end{equation}
where $f_i(t)$ is the portion of the colony state contributed by forager $i$ by time $t$, and $N$ is the number of foragers.

\subsubsection*{Feedback on the individual forager scale}

 Collective food inflow ($\frac{dF}{dt}$), as well as the flow of food through each individual forager ($\frac{df_i}{dt}$) were derived by differentiating the measured colony state and the individual contributions with respect to time (Figs. \ref{fig:global}c, \ref{fig:averageforager}, and Methods, Data Analysis). This revealed that flows of food through each individual forager declined with increasing colony satiation state, $F$,  and were roughly proportional to the available space in the colony, $1-F$:

\begin{equation}
    \forall i, \ \frac{df_i}{dt}\approx m(1-F(t))
    \label{eq. dfdt_i}
\end{equation}
where $m$ is a constant (Fig. \ref{fig:global}c and \ref{fig:averageforager}). This linear relationship holds for each forager, regardless of when she began foraging and is thus incompatible with feed-forward control in which a forager slows down as a function of her own history. Rather, it supports a mechanism by which the colony state feeds back on the food transfer rate of each individual forager.

Breaking down the total inflow of food into individual forager contributions and using equation \ref{eq. dfdt_i}, we obtain:
\begin{equation}
    \frac{dF}{dt}= \sum_{i=1}^{n(t)} \frac{df_i}{dt}\approx n(t)\cdot m(1-F(t))
    \label{eq. dFdt}
\end{equation}
where $n(t)$ is the number of foragers that have begun foraging by time $t$.
This formulation provides simple intuition for the non-monotonicity of food flow as apparent in Fig. \ref{fig:global}c. Specifically, an initial rise in  collective inflow occurred when the number of foragers grew at a rate that overcame the rate of individual flow decay. Once  the number of active foragers stabilized total flow rates declined linearly.

Equation \ref{eq. dFdt} describes a feedback process in which the rate of change in the colony state depends on the colony state itself, and more specifically - on $1-F$, the space left to fill until the colony reaches its target. This is a direct consequence of individual foragers that deliver food at slower rates as the colony fills (Eq. \ref{eq. dfdt_i}). However, the satiation state of the colony is a global factor that is, most likely, not directly available to individual ants. In the next section, we demonstrate how the observed feedback emerges from pairwise trophallactic interactions.

\subsubsection*{Global feedback from local interactions}

The average food flow through a single forager ($\frac{df_i}{dt}$) can be estimated by the product of two macroscopic parameters that may depend on the colony state, $F$: her average interaction rate, $\langle r(F) \rangle$, and the average volume transferred per interaction, $\langle v(F) \rangle$.

\begin{equation}
    \frac{df_i}{dt}\approx\langle r(F) \rangle \cdot \langle v(F) \rangle
    \label{eq. rv}
\end{equation}

\begin{figure}
\centering
    \includegraphics[scale=0.48]{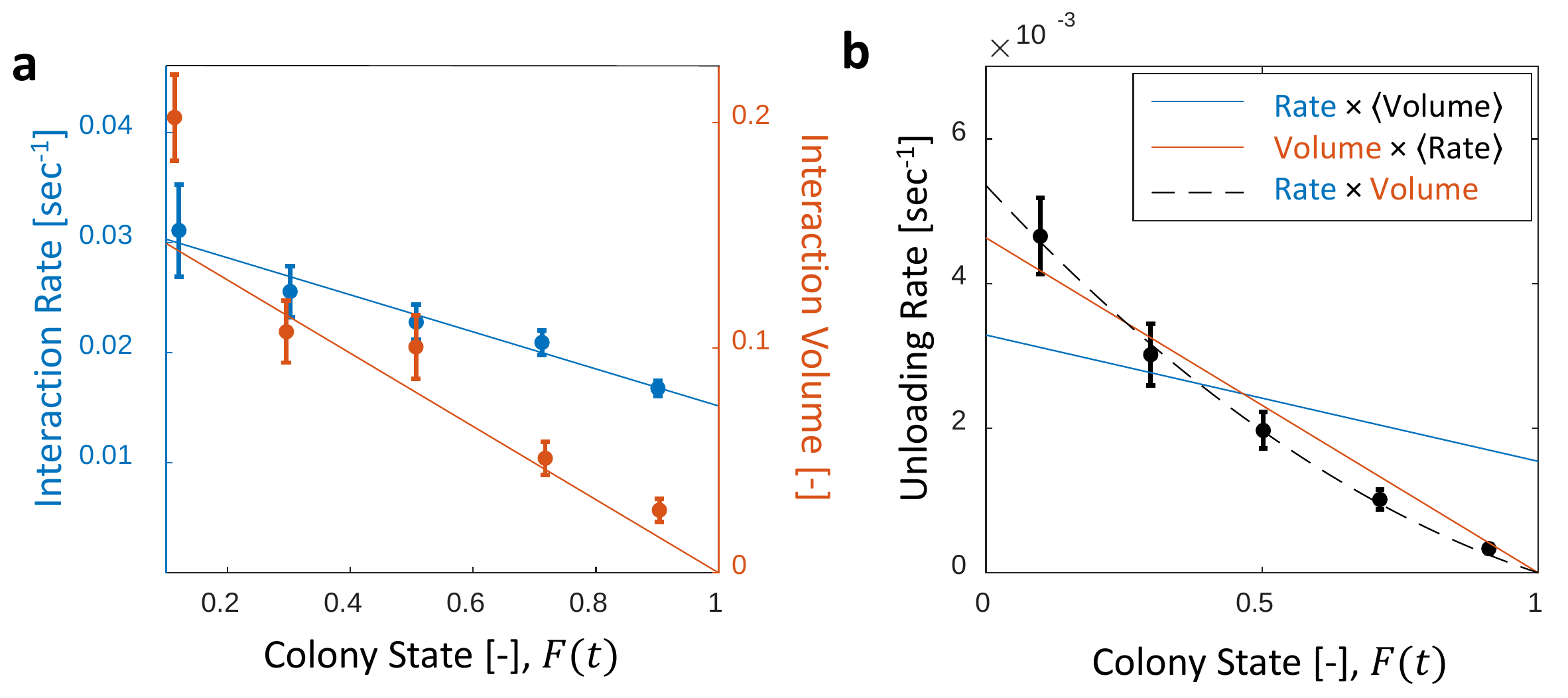}

    \caption{\textbf{Interaction Volume is the Dominant Component of Unloading Rate.}
    \textbf{(a)} Interaction rate (blue) and interaction volumes (red) both decline with increasing colony state. Binned data is presented by mean $\pm$ SEM. Interaction rate was calculated as the inverse of all intervals between interactions. The intervals were binned according to the colony state at which they occurred (n = 49, 79, 101, 170, 240 for bins 1-5, respectively. See Fig. \ref{fig:intrate_sup} for raw data and Figure 2-source data 1 for binning sensitivity analysis). Blue line depicts a linear fit $r(F)=0.032-0.017F$, $R^2=0.96$. Interaction volumes were measured in units of pixel intensities, normalized between experiments (see Methods, Data Analysis), and binned into equally-spaced colony state bins (n = 84, 137, 165, 274, 496 for bins 1-5, respectively. See \ref{fig:intvolume_sup}). Red line represents the predicted relationship between the mean interaction volume and the colony state from Eq. \ref{eq:meanVF}, with $C_0=1.14$ and $\frac{1}{\lambda_0}=0.14$ as obtained from the fit in Fig. \ref{fig:vdist}a. \textbf{(b)} Foragers' unloading rates at each visit in the nest were binned according to colony state (black, mean $\pm$ SEM for each bin, n=26,26,28,39,57 for bins 1-5, respectively).  Mean unloading rate values were fitted by three functions: the blue line represents a model which includes the effect of interaction rates only ($unloading\ rate\propto0.032-0.017F$, function obtained from fit in panel a, $R^2=0.52$), the red line represents a model which includes the effect of interaction volumes only ($unloading\ rate\propto0.2-0.2F$, function obtained from fit in Fig. \ref{fig:intvolume_sup}, $R^2=0.96$), and the black dashed line represents a model that incorporates the combined effects of interaction volumes and interaction rates ($unloading\ rate\propto(0.032-0.071F)\cdot(0.2-0.2F)$, $R^2=0.99$).
    All panels in this figure represent pooled data from all three observation experiments. For raw data see Fig. \ref{fig:twoFits}. Source file is available in  Figure 1—source data 1.}
    \label{fig:rateVolume}
\end{figure}

While both the interaction rate and the interaction volume declined with increasing colony state, the change in the interaction volume was more prominent (Fig. \ref{fig:rateVolume}a, \ref{fig:intrate_sup}, and \ref{fig:intvolume_sup}). In fact, interaction volumes were nearly sufficient to account for the inflow dynamics, while the interaction rate introduced a minor second-order correction (Fig. \ref{fig:rateVolume}b and \ref{fig:twoFits}). Importantly, interaction rates alone did not suffice to account for the inflow dynamics (Fig. \ref{fig:rateVolume}b).

\begin{figure}
\centering
    \includegraphics[scale=0.48]{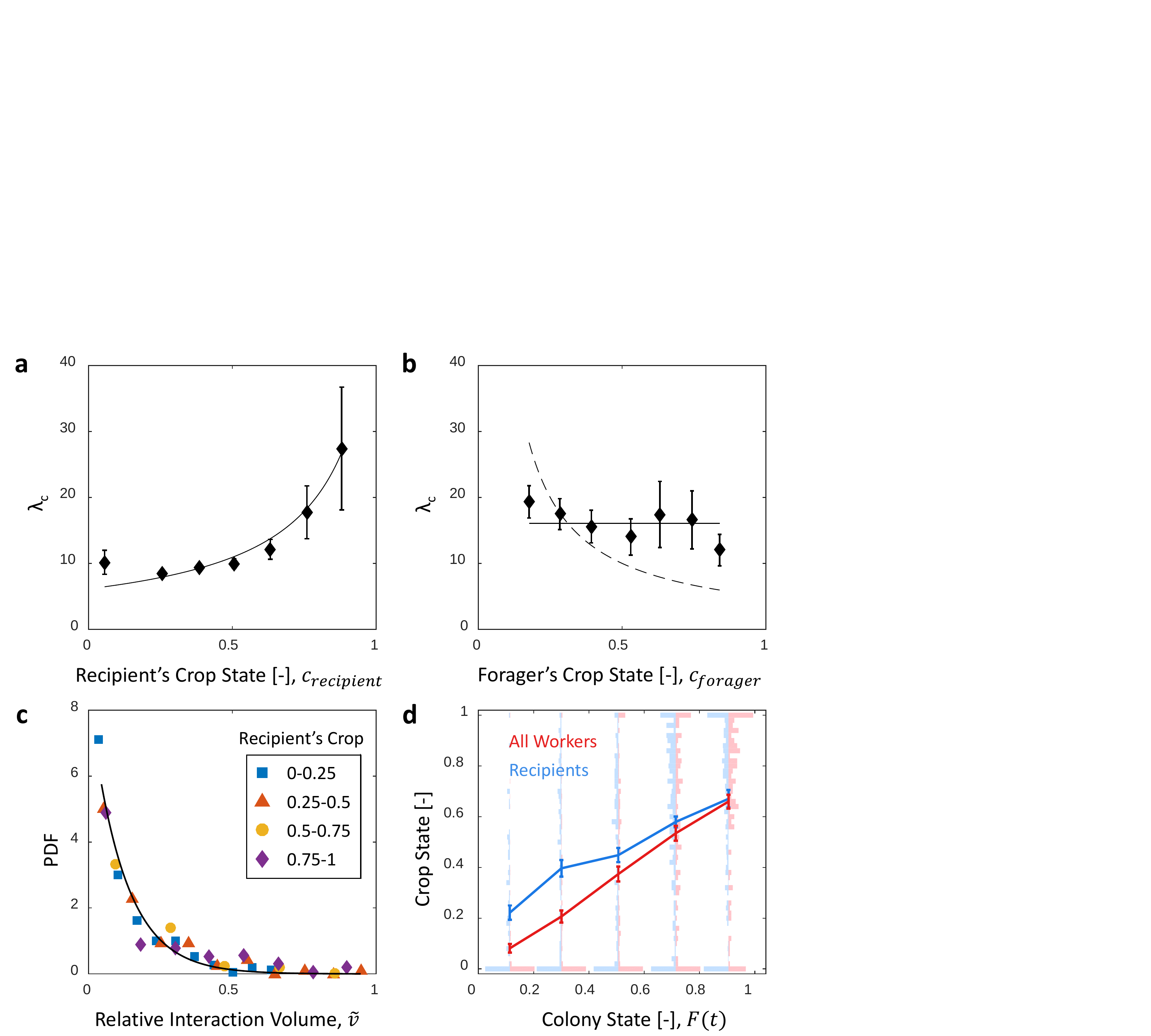}

    \caption{\textbf{Microscopic Food Flow.}
    \textbf{(a)} The distributions of interaction volumes from foragers to non-forager recipients, at seven ranges of \textit{recipient} crop loads, follow an exponential probability density function of the form $p(v|c)=\lambda_c e^{-\lambda_c v}$ (Eq. \ref{eq:pvc}, Fig. \ref{fig:sevenFitsR}). Here the seven rate parameters of the distributions ($\lambda_c$) are plotted as a function of the \textit{recipient's} crop load ($c_{recipient}$). Mean $\pm$ STD of $\lambda_c$ are over different binnings of the histograms to which the exponential distribution function was fit (17 histogram binwidths uniformly covering the range [0.01-0.09]). Curve represents a fit of the function $\lambda_c=\frac{\lambda_0}{C_0-c}$ (Eq. \ref{eq:lambda}): $\lambda_0=7.01$, $C_0=1.14$, $R^2=0.93$.
    \textbf{(b)} Similarly to panel a, the rate parameters of the exponential distributions of interaction volumes were obtained for seven ranges of \textit{forager} crop loads (Fig. \ref{fig:sevenFitsF}) and plotted as a function of the \textit{forager's} crop load, $c_{forager}$. Dashed curve represents a fit of the form $\lambda_c=\frac{const}{c_{forager}}$, similar to Eq. \ref{eq:lambda}, but instead of a fraction of the recipient's empty crop space, $\lambda_c$ is assumed to be a fraction of the forager's crop load. $R^2$ was negative, indicating that this function is no better fit to the data than a constant (solid line).
    \textbf{(c)} The distributions of ``relative interaction volumes", $\tilde{v}$, wherein each interaction volume was normalized to the available space in the receiver's crop ($\tilde{v}=\frac{v}{C_0-c}$). The distributions collapse onto a single exponential function $p(\tilde{v})=\frac{1}{\lambda}e^{-\frac{1}{\lambda}\tilde{v}}$, $\lambda=0.12$, $R^2=0.96$. \textbf{(d)} The crop loads of non-foragers at interactions with foragers (blue, n as in b) compared to crop loads of all non-foragers in the colony (red, n=202 per colony state bin). The distributions of crop loads at each colony state are plotted as a violin plot, and the mean $\pm$ SEM are plotted in solid lines. All panels in this figure represent pooled data from all three observation experiments. Source file is available in the Figure 1—source data 1.}
    \label{fig:vdist}
\end{figure}

Therefore, we  turned to explore the local determinants that affect interaction volumes, under the assumption that interaction volumes are set locally depending on the states of the interacting individuals.
The maximal potential volume of any given interaction is constrained by both the donor's crop load and the available space in the recipient's crop. To inspect the impact of each of these two local factors, we examined the distribution of all interaction volumes ($v$) from foragers to non-forager recipients for different ranges of crop loads, either of the recipient ($c_{recipient}$, Fig. \ref{fig:sevenFitsR}) or of the forager ($c_{forager}$, Fig. \ref{fig:sevenFitsF}). We found that  these distributions all follow an exponential probability density function (PDF) of the form:

\begin{equation}
p(v|c)=\lambda_c e^{-\lambda_c v}
\label{eq:pvc}
\end{equation}
where $p(v|c)$ is the conditional PDF of interaction volumes, $v$, given a crop load $c$, and $c$ is either $c_{forager}$ or $c_{recipient}$. We found that while the recipient's crop load affected the distribution of interaction volumes (Fig. \ref{fig:vdist}a), that of the forager had little effect, if any (Fig. \ref{fig:vdist}b). Specifically, the distribution of interaction volumes scaled with the space left to fill in the recipient's crop but was effectively independent of the forager's crop load (hence, hereafter $c=c_{recipient}$):

\begin{equation}
\lambda_c=\frac{\lambda_0}{C_0-c}
\label{eq:lambda}
\end{equation}
where $\lambda_0=7.01$ and $C_0=1.14$ (Fig. \ref{fig:vdist}a). $C_0$ may be interpreted as the average crop load target to which recipients aim to ultimately fill (expected to be close to $1$).

On average, in an interaction with a forager the recipient receives $\frac{1}{\lambda_0}=0.14$ of the space left to fill in her crop. Consistently, a linear fit relating mean interaction volume to recipient crop load $ \langle v(c) \rangle=ac+b$ (Fig. \ref{fig:volume_receiver}) yields $b\approx -a\approx0.13\approx \frac{1}{\lambda_0}$ . This is to be expected if   $ \langle v(c) \rangle=\frac{1}{\lambda_0}(1-c)$ .
Indeed, normalizing interaction volumes by the total amount of available space in the recipient's crop, $\tilde{v}=\frac{v}{C_0-c}$, we find that all trophallactic volume distributions collapse onto a  single exponential function $p(\tilde{v})=\lambda e^{-\lambda\tilde{v}}$ with $\frac{1}{\lambda}=0.12\approx \frac{1}{\lambda_0}$ (Fig. \ref{fig:vdist}c). Simply put, the volume of an interaction is a random exponentially distributed fraction ($\tilde{v}$) of the available space in the recipient's crop ($C_0-c$).

We can now move forward to express mean interaction volume, $ \langle v  \rangle$, in terms of the colony state. Defining $p(v|F)$ to be the conditional probability for an interaction of volume $v$ when the colony state is $F$, the mean interaction volume (at colony state $F$) can be calculated by:

\begin{equation}
    \langle v(F) \rangle =  \int\limits_v v \cdot p(v|F) dv
    \label{eq:average_v_col}
\end{equation}

In light of our findings that interaction volumes change mainly with respect to the recipient's crop load, we can decompose $p(v|F)= \int\limits_c p(v|c)\cdot p(c|F) dc$ where $p(c|F)$ is the probability density that the recipient will have a crop load of size $c$ at colony state $F$. Eq.\ref{eq:average_v_col} now becomes:

\begin{equation}
    \langle v(F) \rangle = \int\limits_c p(c|F) \int\limits_v v \  p(v|c) \  dv\ dc
    \label{eq:average_v}
\end{equation}
This probability ($p(c|F)$) changed as the colony satiated and individual ants approached their targets. Figure \ref{fig:vdist}d shows that the ants that interact with a forager reliably represent the satiation level of the colony and that the accuracy of this representation increases as the colony satiates. Altogether, substituting the microscopic interaction rule described by equations \ref{eq:pvc} and \ref{eq:lambda} into the global summation described by equation \ref{eq:average_v} demonstrates how the average interaction volume changes in proportion to the empty space in the colony ($1-F$):

\begin{equation}
    \langle v \rangle = \int\limits_c p(c|F) \int\limits_v \frac{\lambda_0}{C_0-c}e^{-\frac{\lambda_0}{C_0-c}v}\ v\ dv \ dc = \int\limits_c p(c|F)\frac{C_0-c}{\lambda_0} \ dc = \frac{1}{\lambda_0}(C_0-\langle c(F) \rangle) = \frac{C_0}{\lambda_0}(1-F)
\label{eq:meanVF}
\end{equation}
where 
the following identities were used:
$C_0 = \frac{ \sum{c_{target}}}{N}$,\ \ \
$F=\frac{\sum{c}}{\sum{c_{target}}}$,\ \ \
$\int p(c)dc=1$ \ \ \ and\ \ \
$\int p(c)cdc=\langle c\rangle$. For each ant, $c_{target}$ signifies her crop load at colony satiation. The value of the multiplicative factor $\frac{C_0}{\lambda_0}\approx 0.16$ stands in agreement with our experimental measurements (Fig. \ref{fig:intvolume_sup}).

The above analysis demonstrates how the global inflow is determined by interaction volumes that are locally controlled by the recipient's crop load, which on average represents the colony state. The forager's crop comes into play in a different aspect of food inflow. Its finite capacity requires foragers to repeatedly leave the nest to reload at the food source in order to supply food to the entire colony. However, leaving the nest encompasses inevitable risks and energetic costs. Therefore, it is interesting to study whether foraging effort is also regulated, and if so, how it is expressed in decisions of individual foragers to leave the nest.

\subsection*{Foraging Effort is Matched to the Colony's Needs}
\begin{figure}

    \centering
    \includegraphics[scale=0.33]{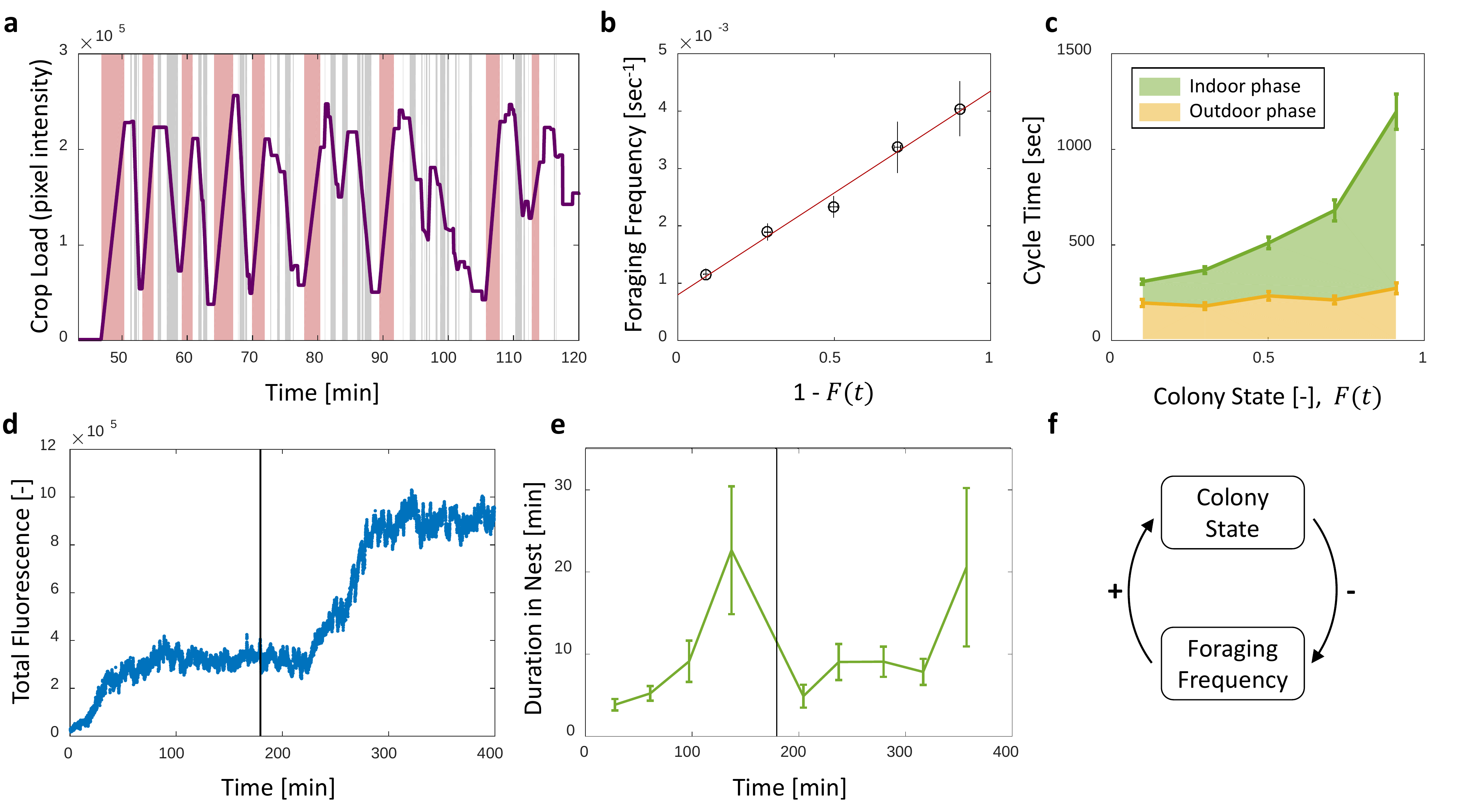}
    \caption{\textbf{Foraging Cycles}. \textbf{(a)} The estimated crop load of a single forager during the first two hours of an experiment. As typical for a forager, her crop load oscillates as she alternates between feeding at the food source (pink areas) and unloading in trophallaxis (gray areas) in continuous back-and-forth trips. \textbf{(b)} Foraging frequency, calculated as the inverse of cycle times (the time interval between two consecutive feeding events of a single forager), grows linearly with the empty space in the colony, $1-F$. Data points and error bars represent means and SEM of cycles. The pooled data from all three observation experiments is grouped into equally-spaced bins of colony state (n=57,39,28,26,26, for bins 1-5, respectively, see Fig. \ref{fig:cycleSI}). A linear fit is presented in red: $y=0.8\ 10^{-3}+3.6\ 10^{-3}(1-F)$, $R^2=0.98$.
    \textbf{(c)} Forager cycle durations are composed of an indoor phase (green) and an outer phase (yellow), the former accounting for most of the rising trend. The pooled data from all three observation experiments was binned and averaged as in panel b (n=26,26,28,39,57, for bins 1-5, respectively, Fig. \ref{fig:cycleSI}).
    \textbf{(d)} Food accumulation in a perturbation experiment. Food rises to an initial plateau, and rises again to a secondary plateau after new hungry ants are introduced (black line). \textbf{(e)} Durations of foragers in the nest in the manipulation experiment described in panel d. Durations grow longer, drop after new hungry ants are introduced (black line), and subsequently rise again. Data points and error bars represent means and standard errors of durations of cycles grouped into time bins (n=28,36,21,14,9,28,27,19,5, for bins 1-9, respectively). Raw data and results from a second replication of the perturbation experiment are presented in Fig. \ref{fig:manipSI}. \textbf{(f)} A schematic representation of the observed negative feedback between the colony state and the foraging frequency. Source file for panels b and c is available in the Figure 4—source data 1. Source file for panels d and e is available in the Figure 4—source data 2.}
    \label{fig:cycles}
\end{figure}

To check whether foragers adjust their activity  to the changing colony state we examined their behavior with respect to the accumulating food in the colony. The behavior of individual foragers was typically of a cyclic nature, as they alternated between two phases: (1) an outdoor phase, in which they filled their crops at the food source, and (2) an indoor phase, in which they distributed their crop contents in trophallaxis to colony members inside the nest (Fig. \ref{fig:cycles}a).

The frequency of these cycles (``foraging frequency") displayed a tight linear relationship with the available space in the colony ($1-F$), demonstrating that foraging effort is matched to the colony's needs (Fig. \ref{fig:cycles}b and \ref{fig:cycleSI}a, $y=0.8\ 10^{-3}+3.6\ 10^{-3}(1-F)$, $R^2=0.98$). Additionally, the increase in cycle times was mainly attributed to the prolonged indoor phase of the cycle, rather than the relatively constant outdoor phase (Fig. \ref{fig:cycles}c and \ref{fig:cycleSI}b, Spearman's correlation test, indoor phase: $r_s=0.77$, $p<0.001$, outdoor phase: $r_s=0.13$, $p=0.08$). This suggests that foraging frequency was regulated by the colony.

To test for a causal effect of colony state on foraging frequency, we elicited an external
perturbation on the colony state. Indeed, in experiments where the colony state was actively dropped by introducing new hungry ants after others had reached satiation (see Methods, Perturbation Experiment), foragers' durations in the nest sharply dropped as well (Fig. \ref{fig:cycles}e and \ref{fig:cycleSI}). This response generated a secondary rise in the amount of food in the colony, relaxing at a new value as durations in the nest gradually lengthened once again (Fig. \ref{fig:cycles}d-e and \ref{fig:manipSI}). These experiments explicitly decoupled colony state from the time that passed since the initial introduction of food, and thus show that the colony state rather than time was the important factor that affects foraging frequency.

Overall, these findings portray the following negative feedback process: foragers raise the colony state by bringing in food, while the colony state, in turn, inhibits their foraging frequency (Fig. \ref{fig:cycles}f). A possible mechanism for this feedback might be that foragers do not exit the nest for their next foraging trip before they fully unload  \cite{traniello1977recruitment,gregson2003partial,buffin2009feeding}. In this case unloading rates directly dictate the foraging frequency. This is consistent with the fact that both unloading rate and the foraging frequency are proportional to the total available space, $1-F$.   We explore this hypothesized mechanism in the next section.

\subsubsection*{Foragers' crop loads upon leaving the nest}

\begin{figure}
    \centering
    \includegraphics[scale=0.485]{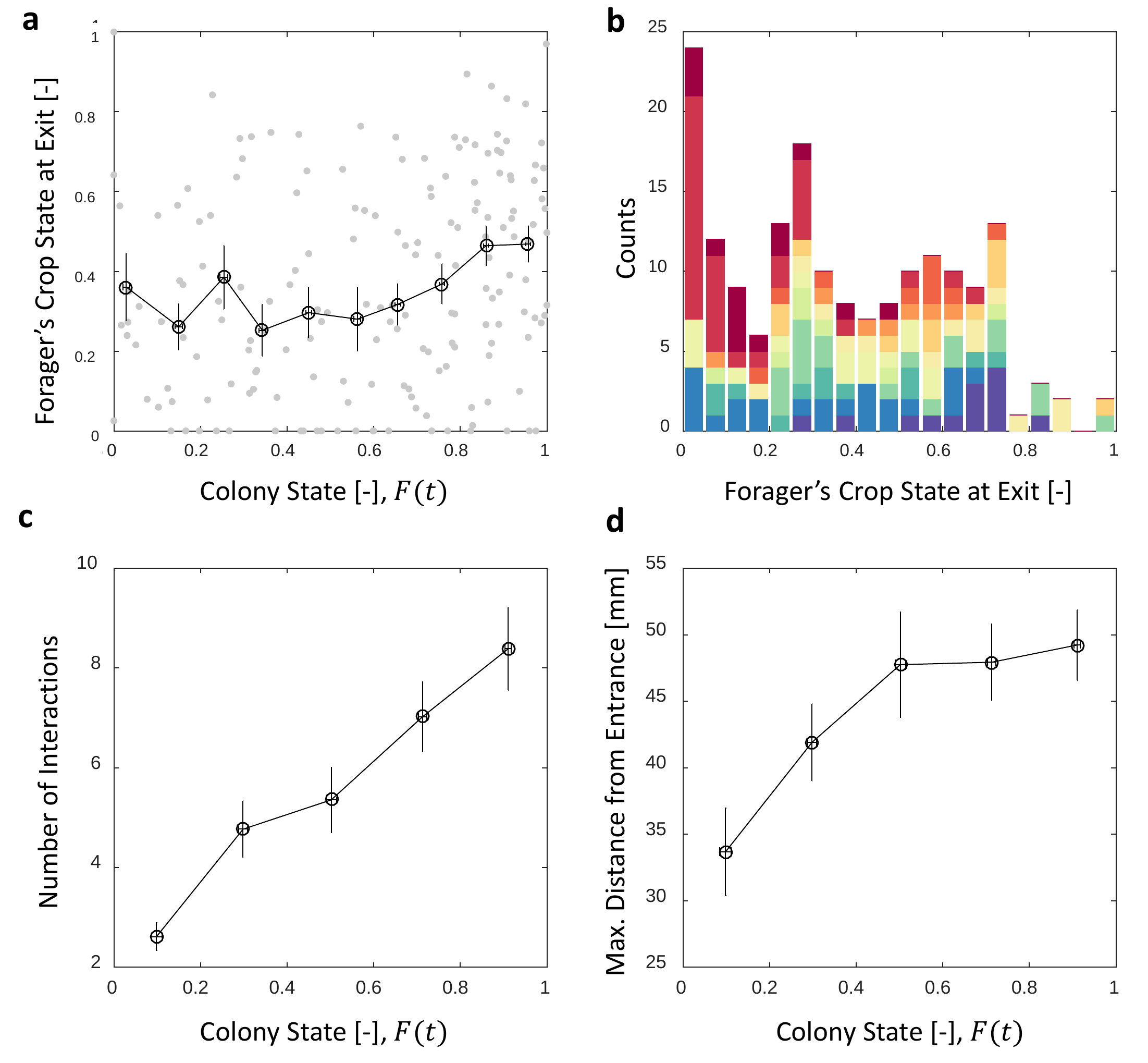}
    \caption{\textbf{Foragers' Crop Loads At Exit.} \textbf{(a)} Foragers' crop loads at the moments of exit were only weakly dependent on colony state (Spearman's correlation test, $r_s=0.24$, $p=0.001$), the average remaining approximately constant with only a slight rise at high colony states. Gray: raw data, black: mean $\pm$ SEM of binned data (n=11,15,13,13,15,13,17,22,27,30 for bins 1-10, respectively).
    \textbf{(b)} The wide distribution of foragers' crop loads at the moments of exit (n=176). Each color represents a different forager, revealing that the distribution of crop loads upon exit is wide within each forager and not because of inter-individual variability.
    \textbf{(c)} The number of interactions a forager has in a single visit to the nest rises as the colony satiates, mean $\pm$ SEM of binned data (n=26,26,28,39,57 for bins 1-5, respectively, Fig. \ref{fig:exitcropSI}a). \textbf{(d)} Foragers reach deeper locations in the nest as the colony satiates, mean $\pm$ SEM of binned data (n as in panel c, Fig. \ref{fig:exitcropSI}b). All panels relate to the pooled data from all three observation experiments. Source file  is available in the Figure 4—source data 1.}
    \label{fig:cropatexit}
\end{figure}

We find that foragers  do not leave the nest only after they have fully unloaded (Fig. \ref{fig:cropatexit}a). However, the average amount of food in their crops when they exit remains nearly constant over different colony states  (Fig. \ref{fig:cropatexit}a, Spearman's correlation test, $r_s=0.24$, $p=0.001$). This constant average is sufficient to produce the observed relation between foraging and unloading rates as specified above.

To maintain a relatively constant average crop state upon exit despite the declining unloading rates, foragers stayed longer in the nest (Fig. \ref{fig:cycles}c) and performed more interactions (Fig. \ref{fig:cropatexit}c, and \ref{fig:exitcropSI}a). They also actively explored deeper into the nest (Fig. \ref{fig:cropatexit}d and \ref{fig:exitcropSI}b). Surprisingly, even though the average crop load with which foragers exit the nest remains constant, this is not because the foragers unload a fixed amount in each visit. Rather, the crop loads with which foragers left the nest were highly variable (Fig. \ref{fig:cropatexit}b). This raises the questions of when foragers decide to exit and how these decisions lead to exits at variable crop contents that, nevertheless, maintain a constant average over different colony states. Do factors other than their own crop load affect their decisions to exit, and more specifically, do the foragers use high-level information regarding the colony state?

\subsubsection*{Forager exit times are determined by their crop load and their unloading rate}
\begin{figure}
    \centering
    \includegraphics[scale=0.485,clip]{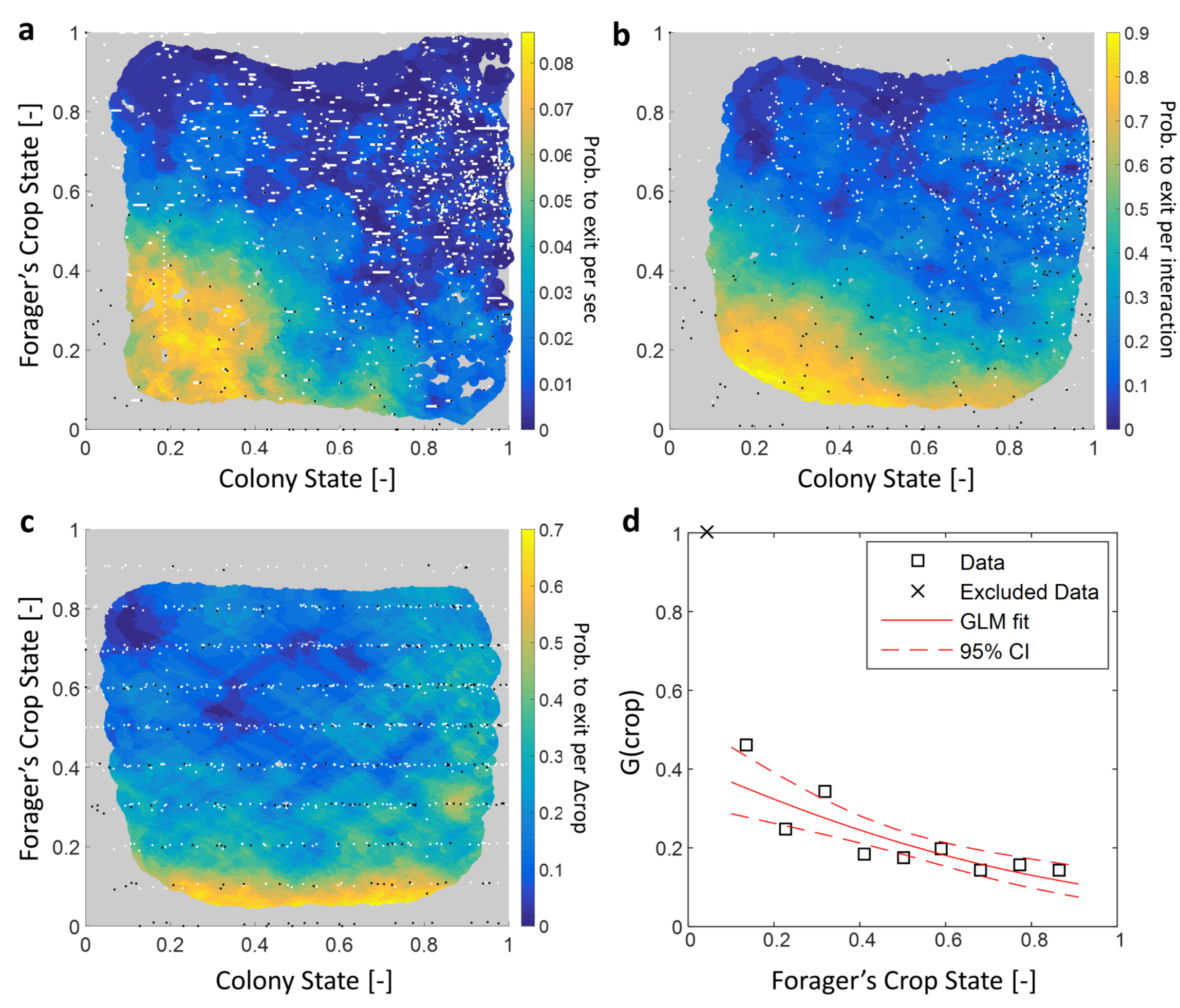}
    \caption{\textbf{Forager Exits.} \textbf{(a-c)} Forager exit probabilities as a function of the colony state and her own crop state. All panels relate to the pooled data from all three observation experiments. Observations are plotted on a 2-dimensional space of the forager's crop state and the colony state, as black and white dots (white - `stay' observations, black - `exits'). An observation was classified as an `exit' if the forager left the nest before the next considered observation. The colored surface represents the estimated local probability to exit on this space, calculated as the fraction of `exits' out of all observations in each bin in the space: the color of each pixel on the surface represents the probability calculated based on the $n$ closest data points, and the pixel's location is the average location of these points (a: $n=300$, b-c: $n=30$). The three panels consider three possible decision rates:  Constant decision rate (a) where all observations, taken every two seconds, were considered to be decision points (excluding observations during trophallaxis).  Decision rate matched to interaction rate (b) where only observations at ends of interactions were considered to be decision points. Decision rate matched to unloading rate (c) where only observations taken each time a forager unloaded $\Delta crop$ food, were considered to be decision points ($\Delta crop$ = 10\% of the forager's capacity). \textbf{(d)} $G(crop)$, a projection of the two-dimensional probability presented in panel c on the forager's crop state axis. Since foragers' crop loads rarely rose in the nest, their lowest crop observation in a visit was generally an `exit', so the calculated probability to exit in the lowest crop interval was 1. To ensure that the crop state played a role beyond this extreme effect, the GLM fit did not include the lowest crop interval. The crop load effect was significant: $\beta_c=-1.93$, $p<0.001$. Source file for panels a and c is available in the Figure 1—source data 2. Source file for panel b is available in the Figure 1—source data 1.}
    \label{fig:P1P2P3}
\end{figure}

To gain insight on the role of individual versus collective information in foragers' decisions to exit, we were interested in a forager's probability to exit the nest as a function of her own crop state ($crop$) and the colony state ($colony$). We first estimated the probability for an individual forager to exit per time unit given her crop load and colony state $R(exit\ |\ crop,colony)$ (hereinafter, $R_{exit}$). We found that this exit rate was strongly dependent on both the forager's crop state and the colony state (Fig. \ref{fig:P1P2P3}a, Table \ref{tab:logreg}).

To understand which information the foragers require to generate this exit pattern we make the simplifying and common assumption that $R_{exit}$ is an outcome of a Markovian decision process \cite{robinson2011simple,sumpter2012modelling}, and  can be treated as a product of two probabilities:

\begin{equation}
    R(exit\ |\ crop,colony)=R(make\ a\ decision)\cdot P(decision=exit\ |\ crop,colony)
    \label{eq.R}
\end{equation}
where $R(make\ a\ decision)$ is the probability of a forager to make a decision within a time unit (hereinafter, $R_{decide}$), and $P(decision=exit\ |\ crop,colony)$ is her probability to decide to exit given her crop load and colony state when a decision is made (hereinafter, $P$).

Since the precise timings of an ant's decisions are beyond our experimental reach, we  replaced $R_{decide}$ by three assumed decision rates: (1) a constant decision rate, (2) a decision rate that is matched to the forager's interaction rate, and (3) a decision rate matched to the forager's unloading rate. Figure \ref{fig:P1P2P3} shows the corresponding $P$ for each decision rate. For a constant decision rate, $P$ is proportional to $R_{exit}$ and depends on both the forager's crop state and the colony state (Fig.  \ref{fig:P1P2P3}a, Table \ref{tab:logreg}). For a decision rate that is matched to the forager's interaction rate, $P$ was calculated by considering only observations at ends of interactions as decision points. In this case, the effect of the colony state on $P$ is present but smaller that the effect of the forager's crop (Fig. \ref{fig:P1P2P3}b, Table \ref{tab:logreg}). Last, for a decision rate matched to the forager's unloading rate, $P$ was calculated by considering observations at fixed intervals of the forager's crop load as decision points. For this case, the effect of the colony state on $P$ approaches zero, such that $P$ 
varies predominantly with the forager's internal crop state (Fig. \ref{fig:P1P2P3}c-d, Table \ref{tab:logreg}).

\begin{table}
    \caption{\textbf{Logistic Fits for a Forager's Probability to Exit ($P$) as a Function of her Crop Load and the Colony State.} A two-dimensional logistic function of the form $P=(1+e^{-(\alpha+\beta\cdot crop+\gamma\cdot colony)})^{-1}$ was fit to each estimated probability to exit from Figure \ref{fig:P1P2P3}a-c. The effect of each factor is reflected in its fitted coefficient. Within each model, effects can be compared to one another because the values of $crop$ and $colony$ lay on the same scale between 0 and 1. In the constant decision rate model, all coefficients were comparable in value, indicating that $crop$ and $colony$ had similar meaningful effects on the probability to exit. In the model where decision rate was matched to interaction rate, the effect of $colony$ was weaker than the effect of $crop$, but both were still meaningful. In the model where decision rate was matched to unloading rate, the effect of $colony$ approached 0 and was very weak compared to the effect of $crop$.}
    \centering
    \begin{tabular}{|c|c|c|c c|c|}\hline

\textbf{Decision Rate Model} & \textbf{Factor} & \textbf{Coefficient} & \multicolumn{2}{|c|}{\textbf{95\% CI}} & $R^2$ \\
\hline
\multirow{3}{4em}{Constant}          & Intercept       & $\alpha=-1.914$      & $-1.921$ & $-1.906$                    & \multirow{3}{2em}{$0.81$}\\
                                     & Colony State    & $\beta=-1.763$       & $-1.778$ & $-1.747$                    &                          \\
                                     & Forager's Crop  & $\gamma=-2.330$      & $-2.347$ & $-2.312$                    &                          \\
\hline
\multirow{3}{8em}{Interaction Rate}  & Intercept       & $\alpha=2.701$       & $2.686$  & $2.716$                     & \multirow{3}{2em}{$0.92$}\\
                                     & Colony State    & $\beta=-2.199$       & $-2.218$ & $-2.181$                    &                          \\
                                     & Forager's Crop  & $\gamma=-5.810$      & $-5.837$ & $-5.783$                    &                          \\
\hline
\multirow{3}{7.5em}{Unloading Rate}    & Intercept       & $\alpha=-0.093$      & $-0.105$ & $-0.081$                  & \multirow{3}{2em}{$0.71$}\\
                                     & Colony State    & $\beta=0.499$        & $0.482$  & $0.515$                     &                          \\
                                     & Forager's Crop  & $\gamma=-3.092$      & $-3.115$ & $-3.069$                    &                          \\
\hline

\end{tabular}
    \label{tab:logreg}
\end{table}

Since the probability to decide to exit, $P$, was effectively independent of the colony state when the rate of decisions, $R_{decide}$, was adjusted to the unloading rate, we learn that the rate of exits $R_{exit}$ can be decomposed into two functions with a clear separation of variables:

\begin{equation}
    R(exit|crop,colony)=U(colony)\cdot G(crop)
    \label{eq. sepVar}
\end{equation}
where $U(colony)$ is a linear function of the unloading rate (Eq. \ref{eq. dfdt_i}) and $G(crop)$ is a function of the crop that does not depend on the colony state (Fig. \ref{fig:P1P2P3}d).
Interpreting this result from the perspective of the individual forager suggests a simple biological mechanism that underlies this separation of variables: In the course of unloading, the forager considers whether to exit or not each time she senses a sufficiently large change in her crop load, and then decides to exit based on her crop load alone. Since the rate at which her crop load changes is mainly affected by the recipients (Fig. \ref{fig:rateVolume} and \ref{fig:vdist}), the rate of her decisions is controlled by the colony ($U(colony)\propto1-F$); once the forager is triggered to make a decision, the decision itself depends on personal information alone ($G(crop)$, Fig. \ref{fig:P1P2P3}d).

\section*{Discussion}

The total flow of food into an ant colony is the sum of many small trophallactic events between foragers and recipient ants. A reliable description of colony level food regulation therefore demands empirical data on the statistics and rules that guide these microscopic events.
In this work, we provide the first quantitative multi-scale account of the dynamics of ant colony satiation that links the global state to local interaction statistics. While our global-scale measurements generally concur with previous studies, our microscopic observations revealed two significant deviations from prevalent individual-level assumptions. First, foragers do not necessarily deliver their entire crop load before exiting the nest. Second, recipients do not fill to their capacity in a single interaction. The next subsections discuss the ways in which our new individual-level observations agree with previous collective level measurements despite the aforementioned deviations. We further discuss how these new findings alter our current understanding of the food intake process.

\subsection*{Regulation of food flow}

On the scale of the entire colony, and in agreement with previous studies \cite{buffin2009feeding,sendova2010emergency}, we find that food accumulation follows logistic dynamics (Fig. \ref{fig:global}b).
Our individual-level measurements confirm that the logistic equation which describes the global dynamics can be interpreted as the product of two intuitive terms (Equation \ref{eq. dFdt}): the number of active foragers times the average unloading rate per forager. As previously speculated \cite{buffin2009feeding}, the initial rise in the global food inflow stems from gradually joining foragers while its subsequent decay  is the result of a negative feedback process wherein colony satiation levels work to decrease the unloading rates of individual foragers (Fig. \ref{fig:global}c).

We traced the mechanisms of this large-scale negative feedback to the immediate experience of individual foragers and specifically to the crop-loads of the ants they interact with. When a forager enters the nest she interacts with a ``representative sample" of recipient ants, i.e. ants whose crop load is, on average, proportional to the total satiation state of the colony (Fig. \ref{fig:vdist}d). Further, in each such interaction the amount of food  transferred is proportional to the available space in the recipient's crop (Fig. \ref{fig:vdist}a,c). Together, these findings imply that unloading rates are determined by the colony and directly proportional to the total empty space in the crops of the entire colony.

Many sets of local rules could have yielded the same average flow, and thus would have been consistent with similar global dynamics.
For example, previous studies have attributed the global negative feedback to the decreasing probability of a forager to encounter an accepting recipient, which delays the time until delivery \cite{sendova2010emergency,seeley1989social,Seeley1994-xw,cassill1999regulation}.
However, due to experimental limitations these studies relied on the implicit assumption that recipients are satisfied by a single interaction, while in fact recipients may very well be partially satiated \cite{huang2003multiple}.
Our measurements on the level of single crops show that recipients are typically partially loaded, and the effect of their crop loads on interaction sizes is more dominant than the minor decrease of interaction rates in generating the collective negative feedback.

Interestingly, partial crop loads do not affect the interaction volume merely by physical limitation: in most interactions the donor does not deliver her entire crop load, nor does the recipient fill up to her capacity. This finding contradicts the prevalent assumption used by those studies that did take partially loaded recipients into account \cite{huang2003multiple,gregson2003partial}. These studies supposed that the amount of food transferred in an interaction is the maximal possible amount, and partial crop loads result from discrepancies between foragers' loads and recipients' capacities. Here we introduce explicit measurements of interaction volumes that reveal that exponentially distributed  interaction volumes lead to partially loaded ants. This volume distribution  concurs with the global feedback as it is scaled to the available space in the recipient's crop. Feedback based on interaction volumes that are not set by physical limitations, rather than ``all-or-none" interactions, potentially allows individual ants to fine-tune their intake and allow for combinations of several sources towards their desired nutritional target \cite{cassill1999regulation}.

\subsection*{Regulation of foraging trips}

Previous studies typically addressed the global feedback between colony state and the collective foraging effort \cite{seeley1989social,Seeley1994-xw,cassill2003rules}. Our work complements this by demonstrating how this feedback acts on individual foraging frequencies (Fig. \ref{fig:cycles}, see also \cite{tenczar2014automated} and \cite{rivera2016quitting}). Furthermore, while previous studies suggested that foragers use local information, such as time delays, interaction rates or number of refused interactions, to infer the colony's needs \cite{seeley1989social,Seeley1994-xw,cassill2003rules,greene2007interaction,gordon1993function}, we propose a mechanism that demonstrates how foragers could adjust their foraging frequency relying on their own crop load alone (Fig. \ref{fig:P1P2P3}c,d) \cite{mayack2013individual}. In brief, foragers could adjust their exit rates to colony needs by modulating their decision rate according to unloading rates, while the decision itself depends on their current crop load alone.

Interestingly, foragers usually do not exit completely empty (Fig. \ref{fig:cropatexit}a,b) as could be intuitively assumed \cite{gregson2003partial,buffin2009feeding}. This provides further evidence that, similar to interaction volumes, foraging  activity is not regulated by pure physical limitations (\textit{i.e.} an empty crop). Rather, we have found that foragers  exit with a wide range of crop loads.
The lack of a well-defined exit threshold entails a potentially wasteful effect in which forager crop loads at exit increase with colony state:
The difficulty of unloading at higher colony states means that foragers spend longer times with a relatively full crop. Since there is a probability to exit at any crop load this may lead to  an upward drift in the crop loads of exiting foragers. Here we show this drift 
is minor (Fig. \ref{fig:cropatexit}a) and propose different options by which this may be achieved.
For example, it could be the case that after each interaction a forager decides whether to exit the nest or, rather, wait for another opportunity to unload.
This decision scheme holds an appealing simplicity as it implies that a forager's decision rate is set externally and not by an internal clock or parameters.  On the other hand, it demands that the forager makes complex decisions that integrate the state of her own crop with colony-level information (Fig. \ref{fig:P1P2P3}b). Another possibility rids the foragers of the need to use colony level information. In this case, foragers effectively modulate  their decisions to exit according to their unloading rates. We suggest a biologically appealing mechanism to achieve this, in which decisions occur at constant crop intervals (Fig. \ref{fig:P1P2P3}c). Generally, any mechanism by which the forager's trigger to leave the nest depends on her unloading rate could yield similar results.

Leaving the nest while partially loaded could hold some benefits: foragers may use the food in their crops as provisions to be consumed in their expedition \cite{rytter2016lunchbox}, waiting for full unloading in the nest may be time-consuming and limit exploration for other food types, and frequent visits to the food source may ensure its exploitation. This raises the question whether there exists an optimal crop load with which foragers should exit the nest, which could potentially depend on factors such as the abundance and quality of the food source, predation risk, and the demand for food in the nest \cite{dornhaus2004information}.

In light of our findings on both the forager's decision to exit and the distribution of interaction volumes, we hypothesize that an internal mechanism based on the mechanical tension of the crop's walls is involved in trophallaxis. Considering the crop as an elastic organ that  stretches as it fills, the relative change in the volume of the recipient's crop may provide a mechanism for the scaling of interaction volumes with available crop space. Additionally, if ants could sense  changes in the tension of their crop walls \cite{stoffolano2013crop}, then this would provide an anatomical basis for a model in which foragers adjust decision rates to unloading rates.

\section*{Methods}

\subsection*{Study Species: \textit{Camponotus sanctus}}
\textit{Camponotus sanctus} are omnivorous ants that are presumed to naturally live in monogynous colonies of tens to hundreds of individuals (projecting from \textit{Camponotus socius}, \cite{tschinkel2005nest}), distributed from the near East to Iran and Afghanistan \cite{ionescu2009annotated}. Workers of this species are relatively large (0.8-1.6 cm) and characterized by translucent gasters, rendering them suitable for both barcode labeling and crop imaging. Our experiments were conducted on lab colonies of 50-100 workers, reared from single queens that were collected during nuptial flights in Neve Shalom and Rehovot, Israel. Table \ref{tab:colonyinfo} contains further details on each experimental colony.

\begin{table}
    \caption{\textbf{Experimental Colonies.} }
    \centering
    \begin{tabular}{|c|c|c|c|c|c|}\hline

\textbf{Experiment Type}          & \textbf{Colony} & \textbf{\# Ants} & \textbf{Starvation Period} & \textbf{\# Major Workers} & \textbf{\# Foragers} \\\hline

\multirow{3}{5em}{Observation}    & A               & 100              & 3 weeks                    &  1                        &  5      \\
                                  & B               & 62               & 5 weeks                    &  1                        &  3      \\
                                  & C               & 53               & 4 weeks                    &  2                        &  4      \\
\hline
\multirow{2}{5em}{Manipulation}   & M1              & 69               & 3 weeks                    &  1                        &  23     \\
                                  & M2              & 95               & 3 weeks                    &  2                        &  16     \\
\hline

\end{tabular}
    \label{tab:colonyinfo}
\end{table}

\subsection*{Tracking Fluorescent Food in Individually Tagged Ants}

\subsubsection*{Experimental setup}
Fluorescent food imaging and 2D barcode identification  (BugTag, Robiotec) were used to obtain a live visualization of the food flow through colonies of individually tagged ants. See \cite{greenwald2015ant} for a detailed description of the experimental setup. In short, an artificial nest was placed on a glass platform positioned between two cameras. A camera below the nest filmed through the platform, capturing the fluorescence emitted from the food inside the translucent ants. Meanwhile, a camera above the nest filmed through its infrared shelter, capturing the barcodes on the ants' thoraxes, allowing identification of single ants inside the nest. Together, footages from both cameras enabled the association between each individual ant and her food load, throughout time and across trophallactic events. The two cameras were synchronously triggered at a fixed frame rate, (here 0.5 Hz., except for colony B which was recorded at 1 Hz.). We chose a temporal resolution that is sufficient to capture events of 2 seconds since shorter interactions barely involve food exchange \cite{greenwald2015ant}.

\subsubsection*{Image processing}
Top camera images were used to extract ant identities, coordinates and orientations using the BugTag software (Robiotec). Bottom camera images were used to detect fluorescence with a pixel intensity threshold, using the openCV library in Python.
Gasters of fed ants appeared as bright “blobs" and thus passed the image threshold (for details, see \cite{greenwald2015ant}).

In order to associate between the identity of an ant and her appropriate blob, the image from the upper camera was transformed to align with the fluorescent image. Then, for each identified tag, a small area extended from the back of the tag toward the ant's abdomen was crossed with the thresholded fluorescent image. If a blob intercepted this area, it was assigned to the tag's identity.

Thus, for each experiment a database was obtained, which included for every frame the coordinates, orientation, and measured fluorescence (in arbitrary units of pixel intensity) of each identified ant.

\subsubsection*{Interaction identification and crop load estimation}
Even though the fluorescence emitted from an ant's crop is reasonably indicative of the food volume, it is a noisy measurement mainly due to her highly variable postures. Therefore, assuming that an ant's crop content remains constant during the intervals between trophallactic events, it is best evaluated as the maximal fluorescence measurement acquired in each such interval \cite{greenwald2015ant}.

In order to precisely consider the relevant intervals for this estimation, the trophallactic interactions were manually identified from the video. Interactions were classified as trophallactic events whenever the mandibles of the participating ants came in contact and the mandibles of at least one of the ants were open. For forager ants, another situation in which their crop loads may change is when they directly feed from the food source. These feedings were also manually identified from the video, as times when a forager's open mandibles touched the food source.

Ultimately, for each ant we obtained a “timeline", describing at every instance whether she was engaged in trophallaxis (and if so, with whom), whether she was directly feeding from the food source, and the estimated food load in her crop. Figure \ref{fig:cycles}a depicts an example of such individual-level data.

\subsection*{Observation Experiment: Monitoring Food Flow as Hungry Colonies Gradually Satiate}
Following a food-deprivation period of 3-5 weeks, ant colonies (queen, workers and brood) were manually barcoded and introduced to the experimental nest for an acclimatization period of at least 4 hours. The nest consisted of an IR-sheltered chamber
(\texttt{\char`\~}100 cm\(^2\)), neighboring an open area which served as a yard (fig. \ref{fig:global}a). After the acclimatization period, the two cameras synchronously started to record. After 30 minutes, the fluorescent food (sucrose [80 g/l], Rhodamine B [0.08 g/l]) was introduced to the nest yard \textit{ad libitum}, and the recording proceeded for at least 4 more hours - a duration sufficient for the colony to reach its desired food volume intake (Fig. \ref{fig:global}b and \ref{fig:globalSI}).

Overall, we analyzed data from 3 such experiments, that included 12 foragers, who fed from the food source 139 times, and were engaged in 1227 trophallactic interactions.

\subsection*{Perturbation Experiment: Introducing Hungry Ants after Initial Satiation}

To manipulatively examine the role of the colony's satiety in the control of food inflow,
we characterized the system's response to a perturbation in the colony’s satiety level. This experiment was conducted as the observation experiment described above, except that it consisted of two phases:

\textit{Phase 1}: The starved colony was segregated between two equally-sized chambers - one with access to the nest yard, and the other blocked behind a removable perspex wall. Thus, when the fluorescent food was introduced to the nest yard, only the ants with access to the yard gradually satiated while the others in the blocked chamber remained hungry. We reasoned that if foragers react to the colony's satiety through their experience in the nest, they would perceive saturation of the accessible chamber as saturation of the colony, as they could only interact with ants of the accessible chamber.

\textit{Phase 2}: After the first chamber satiated, we introduced the hungry ants of the blocked chamber by removing the wall, effectively dropping the perceived satiety level of the colony at once. Recording then proceeded for at least 90 more minutes, sufficient for the colony to reach secondary satiation (Fig. \ref{fig:cycles}d and \ref{fig:manipSI}).

\medskip
\noindent
\textbf{Segregating the colony into two chambers}.
In order to avoid artificial biases in the chambers’ populations, ants were initially introduced to the nest without the wall to freely settle within it. Only after a habituation period of at least 4 hours, the wall was gently inserted to divide the ants, that were then left to habituate for at least one more hour before recording started. The blocked chamber included the queen and brood in both perturbation colonies, and the number of ants in the accessible chamber was 33 and 31 in colonies M1 and M2, respectively.

\medskip
\noindent
\textbf{Time of wall removal}.
Satiation of the first chamber was identified with semi-online approximative image analysis of the videos from the fluorescence camera, by summing the pixel intensities of each frame, which rose as food accumulated. Satiation was determined when this fluorescent signal ceased to rise for at least 1 hour, serving as our cue to remove the wall.

Overall, we analyzed data from 2 perturbation experiments, on colonies of 69 and 95 ants, including 23 and 16 foragers, respectively.

\subsection*{Data Analysis}
All data obtained after crop load estimation was analyzed using Matlab software. Four data files are available with this manuscript.

\subsubsection*{Foragers}
Each experiment consisted of a few individuals who performed consistent foraging cycles between the food source and the nest. Those ants were considered as ``foragers". Some other individuals were occasionally observed at the food source but clearly did not display such foraging cycles. To our purposes they were not considered as foragers. These ants visited the food source no more than 4 times, while consistent foragers performed an average of 15.67 cycles and no less than 8. The data presented here is from the first return of a forger to the nest from the food source until the end of the experiment.

\subsubsection*{Food flow}
The total accumulated food was calculated as the sum of all interaction volumes between foragers and non-foragers. The volume of an interaction was taken to be positive when food was transferred from the forager and negative when it was transferred to the forager. Each forager's contribution is the sum of her own interaction volumes. Although food accumulated through discrete local events, we were mostly interested in the average dynamics of food flow, which are convenient to describe in a continuous manner.
Therefore, the accumulated food was first smoothed with a moving average with a time window large enough to include several trophallactic events (2000 seconds). This window size was chosen by plotting the smoothed data on top of the raw data and assuring that small fluctuations were smoothed while the general shape was maintained. Food inflow was derived by differentiating the smoothed data. Since differentiation is a process highly sensitive to local noise, we differentiated the smoothed accumulated food with a window of 200-500 seconds, depending on the fluctuations of the experiment. This window size was chosen by verifying that the sum of the obtained inflow is indeed sufficiently close to the raw data of accumulated food.

\subsubsection*{Food load normalization}
\begin{figure}
    \centering
    \includegraphics[scale=0.3]{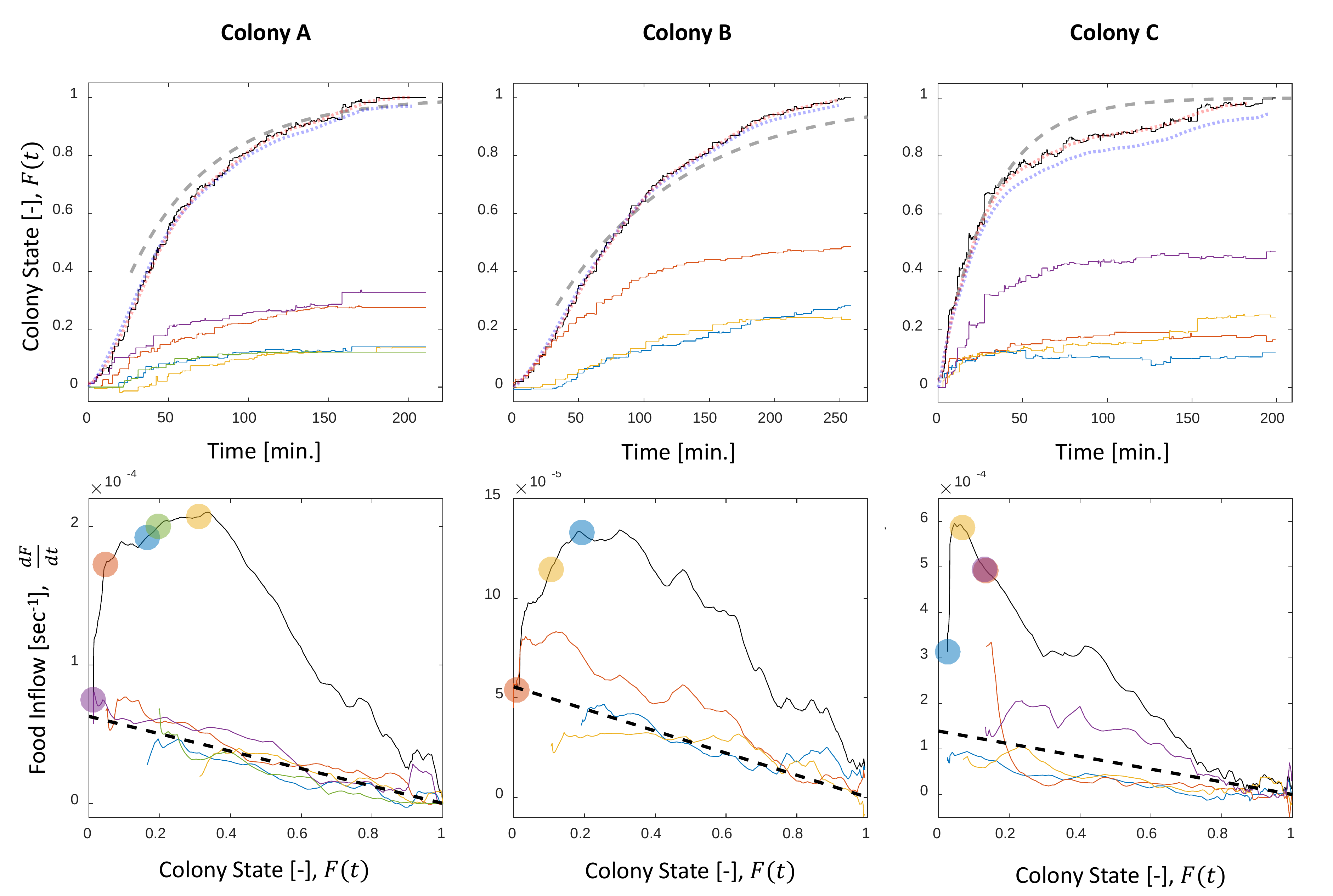}
    \caption{\textbf{Figure \ref{fig:global} - figure supplement 1: Food Accumulation Dynamics in all Three Experimental Colonies.} Food accumulation (top row) and food inflow (bottom row) of each experimental colony (columns). The results of Colony A are presented in Fig. \ref{fig:global} in the main text, where a detailed explanation can be found in the caption. The smoothed data of accumulated food, which is referred to as ``colony state" in all other analyses, and from which the food inflow was derived (see Methods, Food flow), is presented in the red dotted curve. The blue dotted curve is the integral of the obtained food flow. This was plotted to validate the method of differentiation that was used to derive the food inflow (see Methods, Food flow).}
    \label{fig:globalSI}
\end{figure}

\begin{figure}
    \centering
    \includegraphics[scale=0.48]{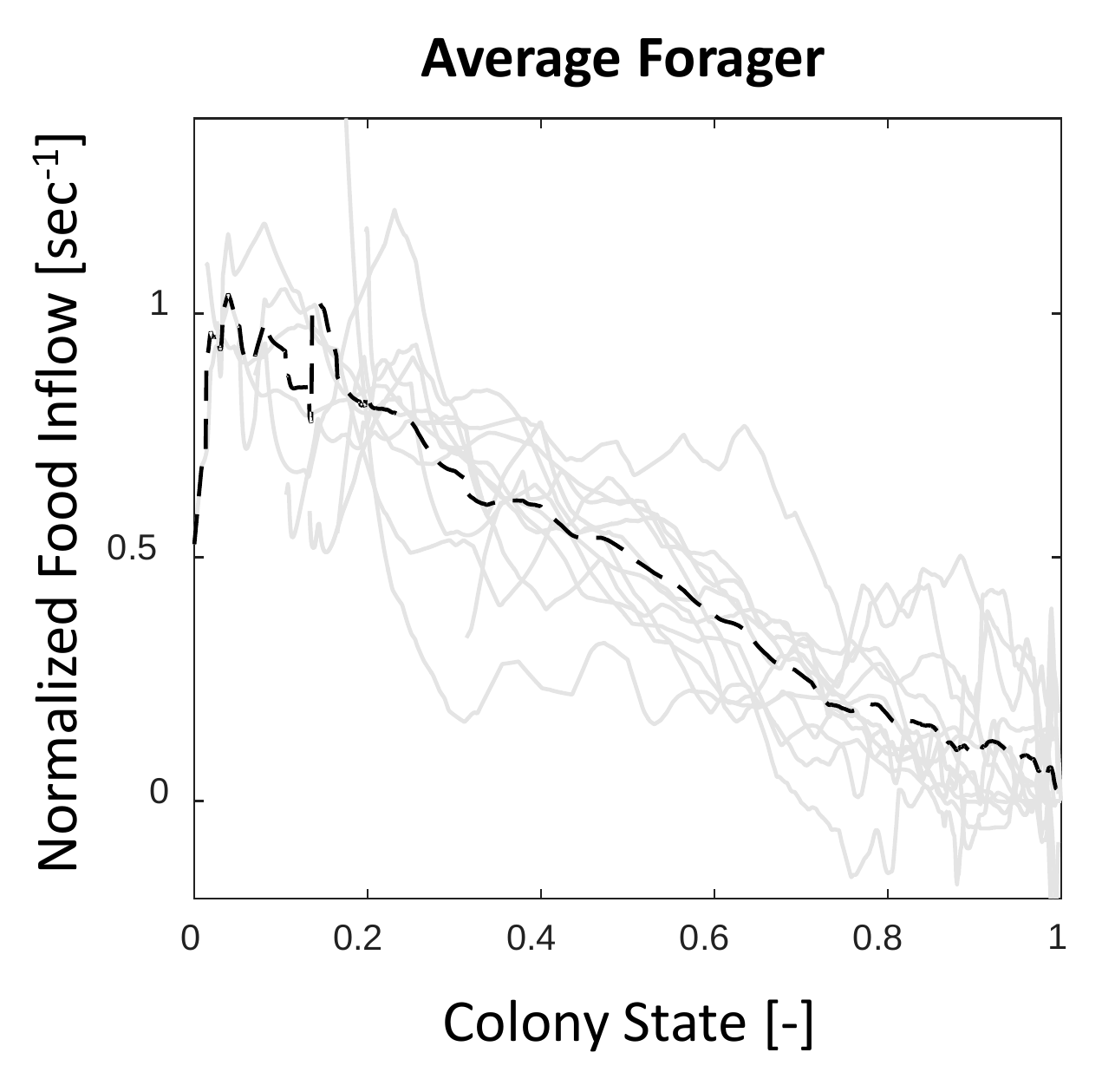}
    \caption{\textbf{Figure \ref{fig:global} - figure supplement 2: Food Flows through Individual Foragers and their Average.} All food flows through individual foragers from all three observation experiments (gray solid lines) were normalized such that $\frac{df_i}{dt}(t=0)=1$, to account for scale differences between experiments. The average of all normalized flows (black dashed line) is highly linear with the colony state. Most foragers, despite starting to forage at different times, do not greatly deviate from this straight line.}
    \label{fig:averageforager}
\end{figure}

\begin{figure}
    \centering
    \includegraphics[scale=0.55]{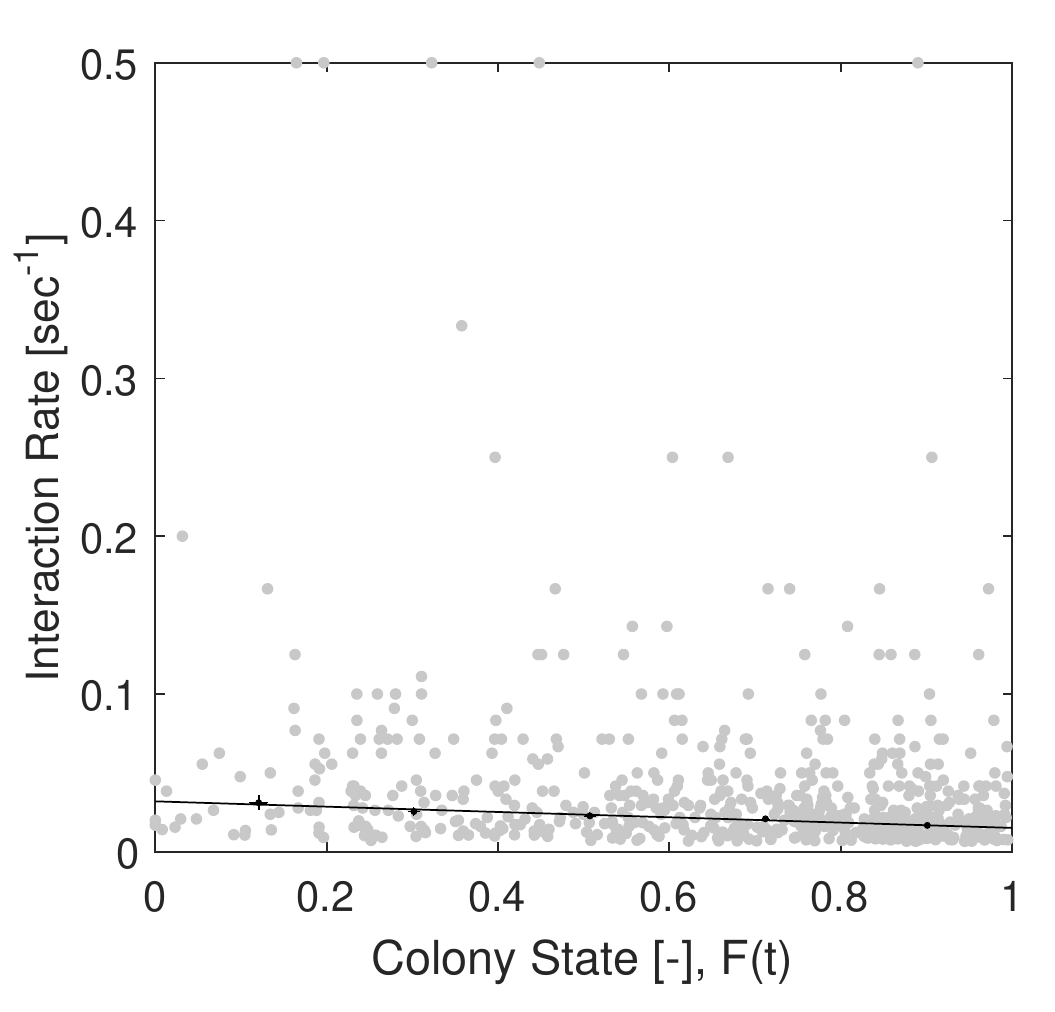}
    \caption{\textbf{Figure \ref{fig:rateVolume} - figure supplement 1: Raw (gray) and binned (black) data of interaction rates}. Each data point is the inverse of an interval between two consecutive interactions of a single forager. Binned data and linear fit are as presented in Fig. \ref{fig:rateVolume}a. The figure relates to the pooled data from all three observation experiments.}
    \label{fig:intrate_sup}
\end{figure}

\begin{figure}
    \centering
    \includegraphics[scale=0.57]{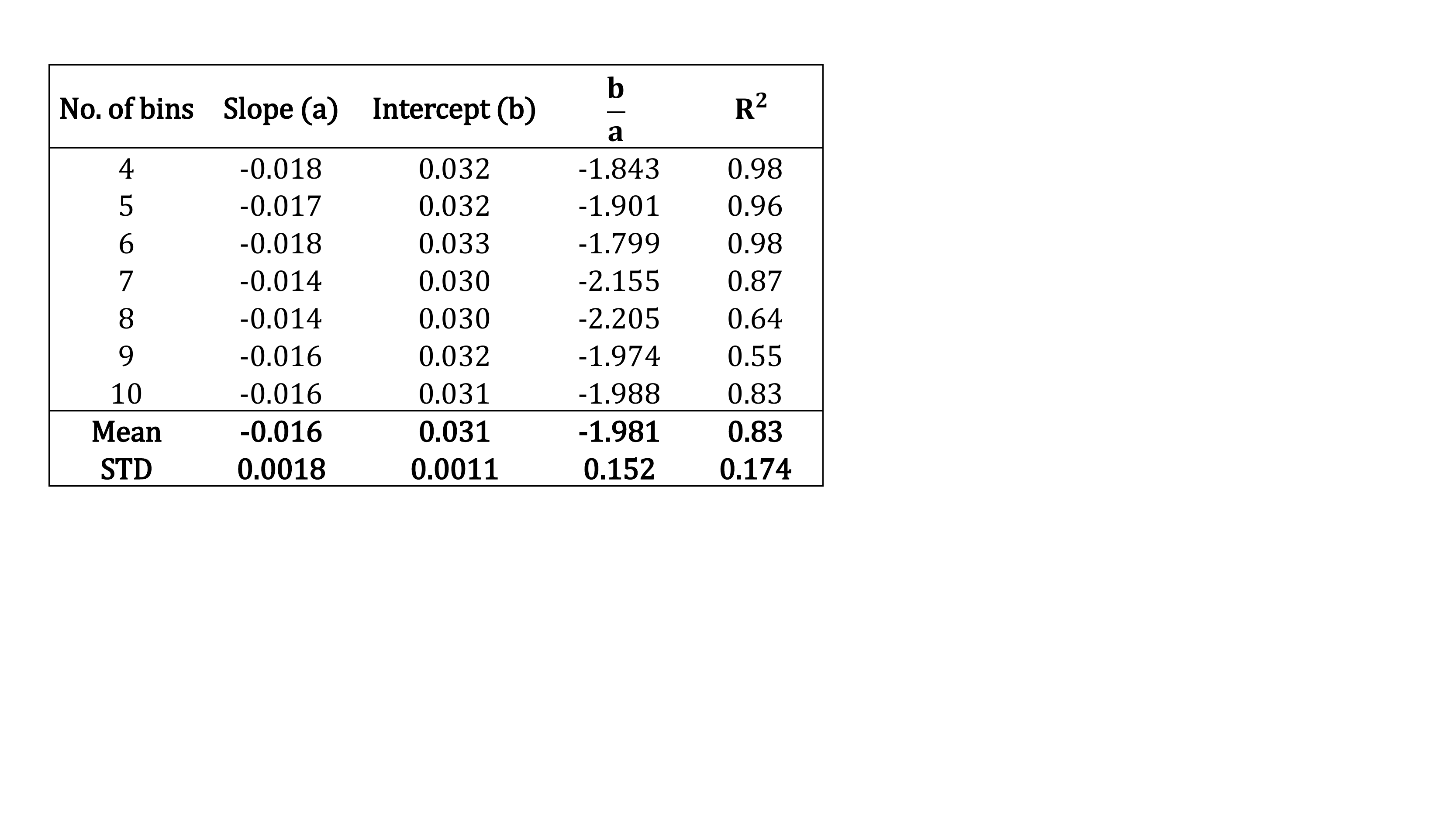}
    \caption{\textbf{Figure \ref{fig:rateVolume} - figure supplement 2: Raw data pooled from all three observation experiments (gray) and binned (black) data of interaction volumes}. Negative values represent interactions in which the forager was the receiver. Binned data and linear fit are as presented in Fig. \ref{fig:rateVolume}a. Linear fits to the raw data (red, $y=0.185-0.178x$, $R^2=0.10$), and to the binned data (blue, $y=0.200-0.201x$, $R^2=0.90$), are similar to each other and close to the prediction (black, $y=0.163-0.163x$, $R^2=0.73$).}
    \label{fig:intvolume_sup}
\end{figure}

\begin{figure}
    \centering

    \includegraphics[scale=0.58]{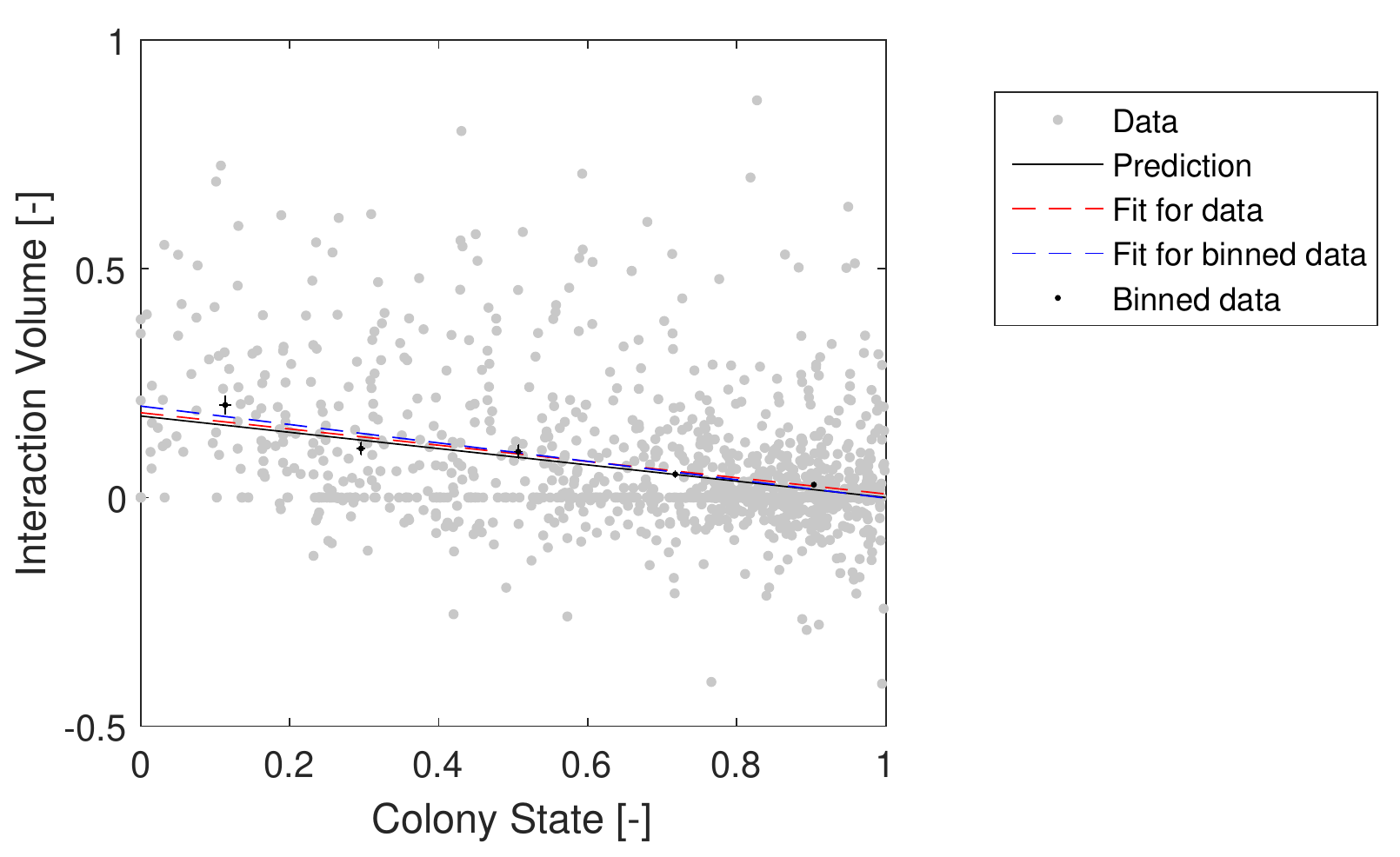}
    \caption{\textbf{Figure \ref{fig:rateVolume} - figure supplement 3: Raw (gray) and binned (black) data of unloading rates.} Each data point is the unloading rate of a forager in a single visit in the nest, calculated as the amount of food she transferred in that visit (in normalized units, see Methods: Data Analysis), divided by the duration of that visit. Binned data and fits are as presented in Fig. \ref{fig:rateVolume}b. The figure relates to the pooled data from all three observation experiments.}
    \label{fig:twoFits}
\end{figure}

\begin{figure}
    \centering

    \includegraphics[scale=0.4]{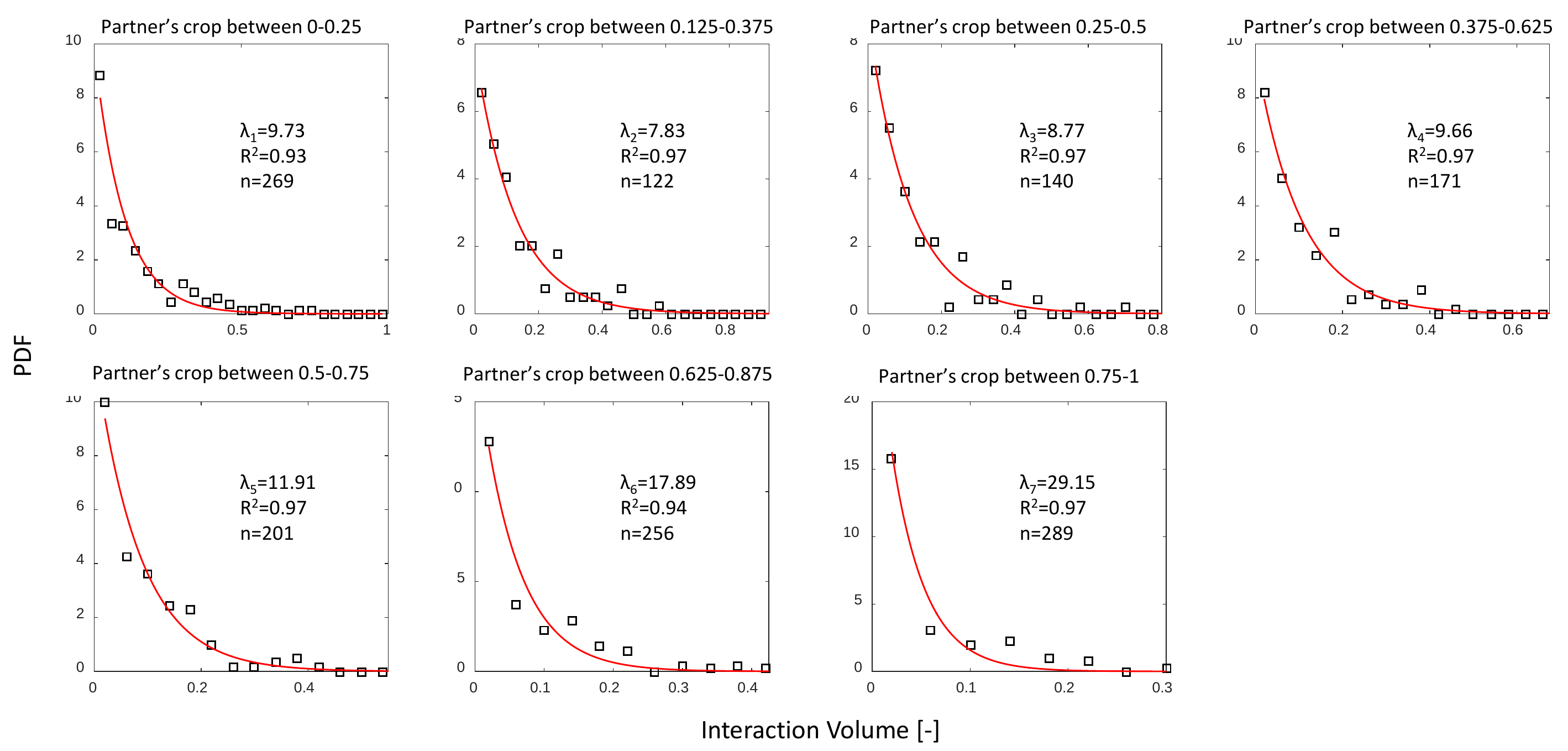}
    \caption{\textbf{Figure \ref{fig:vdist} - figure supplement 1: Interaction Volume Distributions Given Recipient Crop Load.} Each plot depicts a PDF of interaction volumes for a different range of recipient crop loads (indicated in the titles). Data was extracted from pooled data from all three observation experiments. All distributions were fit with exponentials as specified in the text (red curves). The fitted coefficients, $R^2$ of each fit, and sample sizes are indicated on the plots. See Methods section for details on units of food volume. To ensure that this procedure was not sensitive to histogram binwidths, it was performed on a range of binwidths as specified in the caption of Fig. \ref{fig:vdist}a. Here we show the results for a binwidth of 0.04.}
    \label{fig:sevenFitsR}
\end{figure}

\begin{figure}
    \centering

    \includegraphics[scale=0.4]{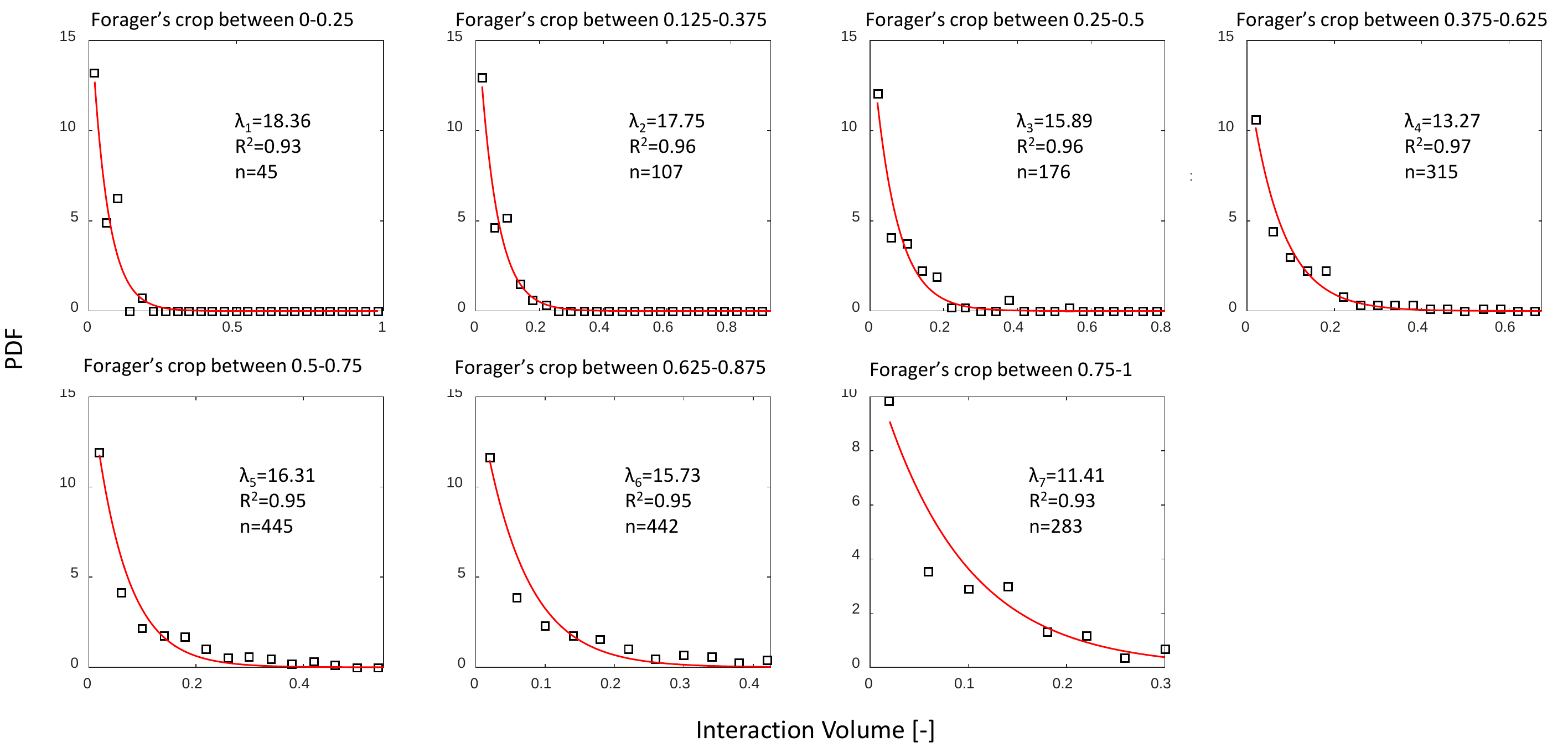}
    \caption{\textbf{Figure \ref{fig:vdist} - figure supplement 2: Interaction Volume Distributions Given Forager's Crop Load.} Each plot depicts a PDF of interaction volumes for a different range of forager crop loads (indicated in the titles).  Data was extracted from  all three observation experiments. All distributions were fit with exponentials as specified in the text (red curves). The fitted coefficients, $R^2$ of each fit, and sample sizes are indicated on the plots. See Methods section for details on units of food volume. To ensure that this procedure was not sensitive to histogram binwidths, it was performed on a range of binwidths as specified in the caption of Fig. \ref{fig:vdist}a. Here we show the results for a binwidth of 0.04.}
    \label{fig:sevenFitsF}
\end{figure}

\begin{figure}
    \centering
    \includegraphics[scale=0.5]{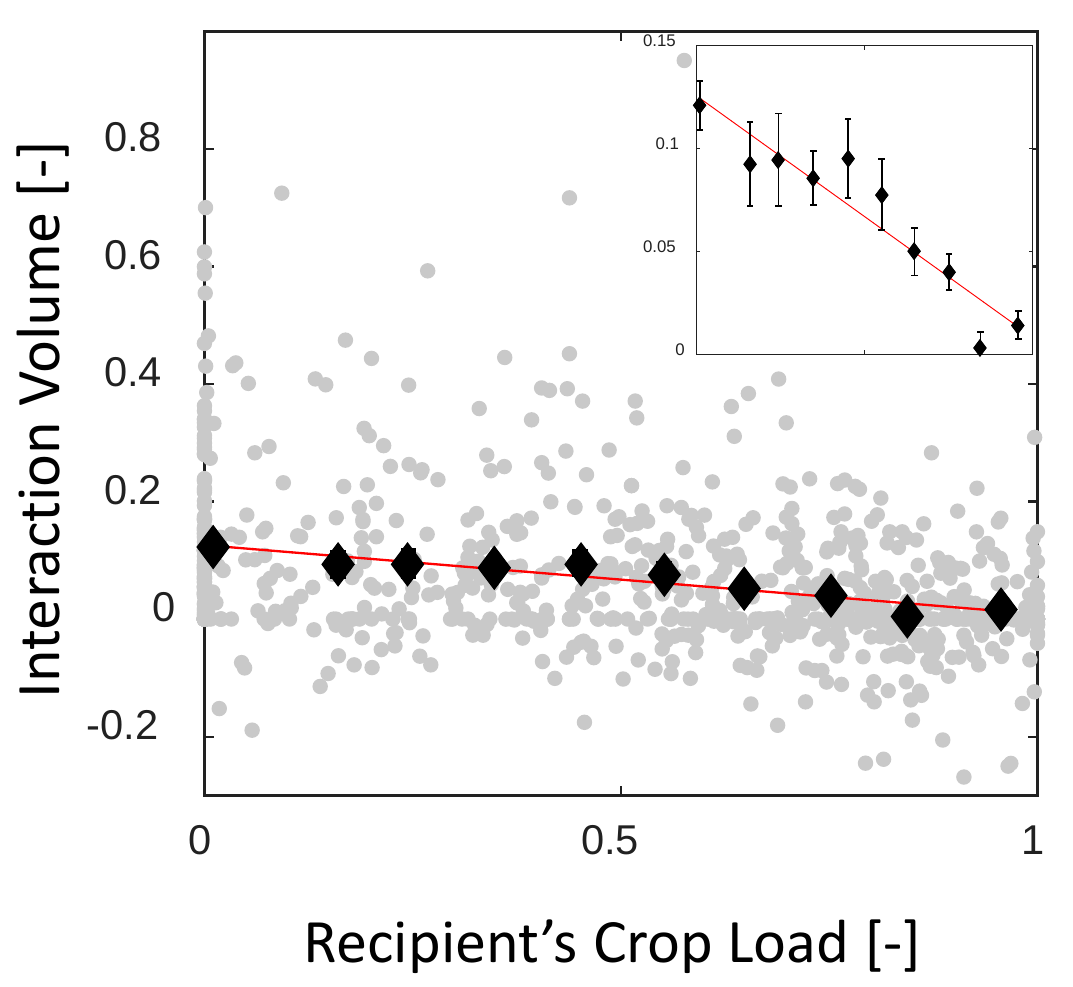}
    \caption{\textbf{Figure \ref{fig:vdist} - figure supplement 3: Mean Interaction Volumes as a Function of the Recipient's Crop Load.} Interaction volumes (gray, pooled data from all three observation experiments) were binned according to the recipient's crop load (black, mean $\pm$ SEM, n=172,44,43,59,64,70,85,100,112,123 for bins 1-10, respectively) and fit with a function of the form $y=ax+b$ (red). Binned data and fit are magnified in the inset. The same fit was obtained for 5-10 bins, resulting in $a=-0.12$ and $b=0.13$ with $R^2=0.93,0.94,0.94,0.94,0.9,0.88$, respectively.}
    \label{fig:volume_receiver}
\end{figure}

\begin{figure}
    \centering
    \includegraphics[scale=0.5]{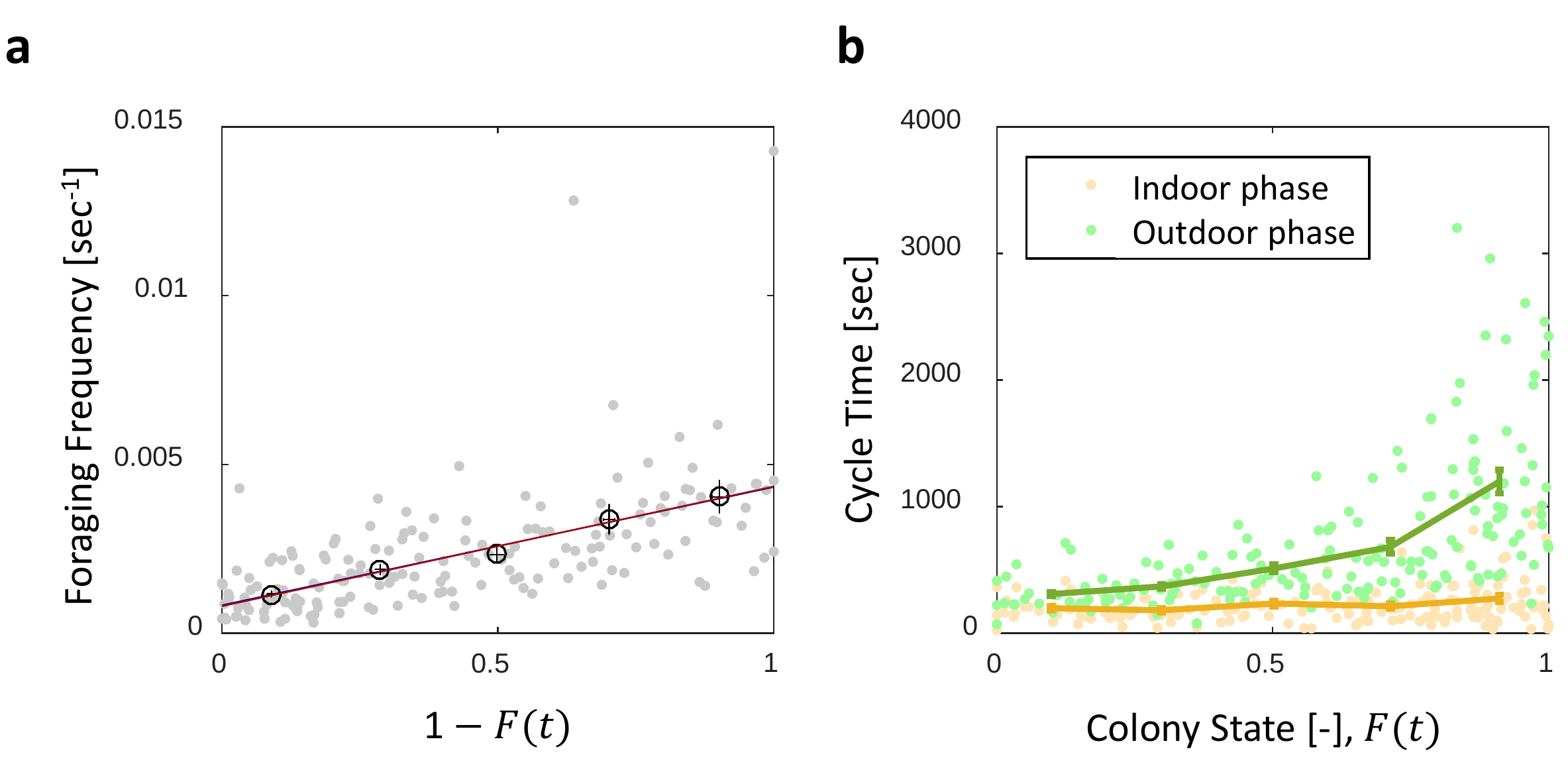}
    \caption{\textbf{Figure \ref{fig:cycles} - figure supplement 1: Foraging Cycle Times. Both panels relate to the pooled data from all three observation experiments}. \textbf{(a)} Foraging frequency, calculated as the inverse of cycle times (the time interval between two consecutive feeding events of a single forager) grows linearly with $1-F$. Raw data (gray) was binned into equally-spaced bins of colony state (n=57,39,28,26,26, for bins 1-5, respectively, in black mean $\pm$ SEM.) Linear fits to the raw data (red) and the binned data (blue, hidden behind the red) yield similar lines: $y=0.8\ 10^{-3}+3.6\ 10^{-3}(1-F)$, $R^2=0.35,0.98$, respectively.
    \textbf{(b)} Forager cycle durations are composed of an indoor phase (green) and an outer phase (yellow), the former accounting for most of the rising trend. Data was binned and averaged as in panel a (n=26,26,28,39,57, for bins 1-5, respectively).}
    \label{fig:cycleSI}
\end{figure}

\begin{figure}
    \centering
    \includegraphics[scale=0.3]{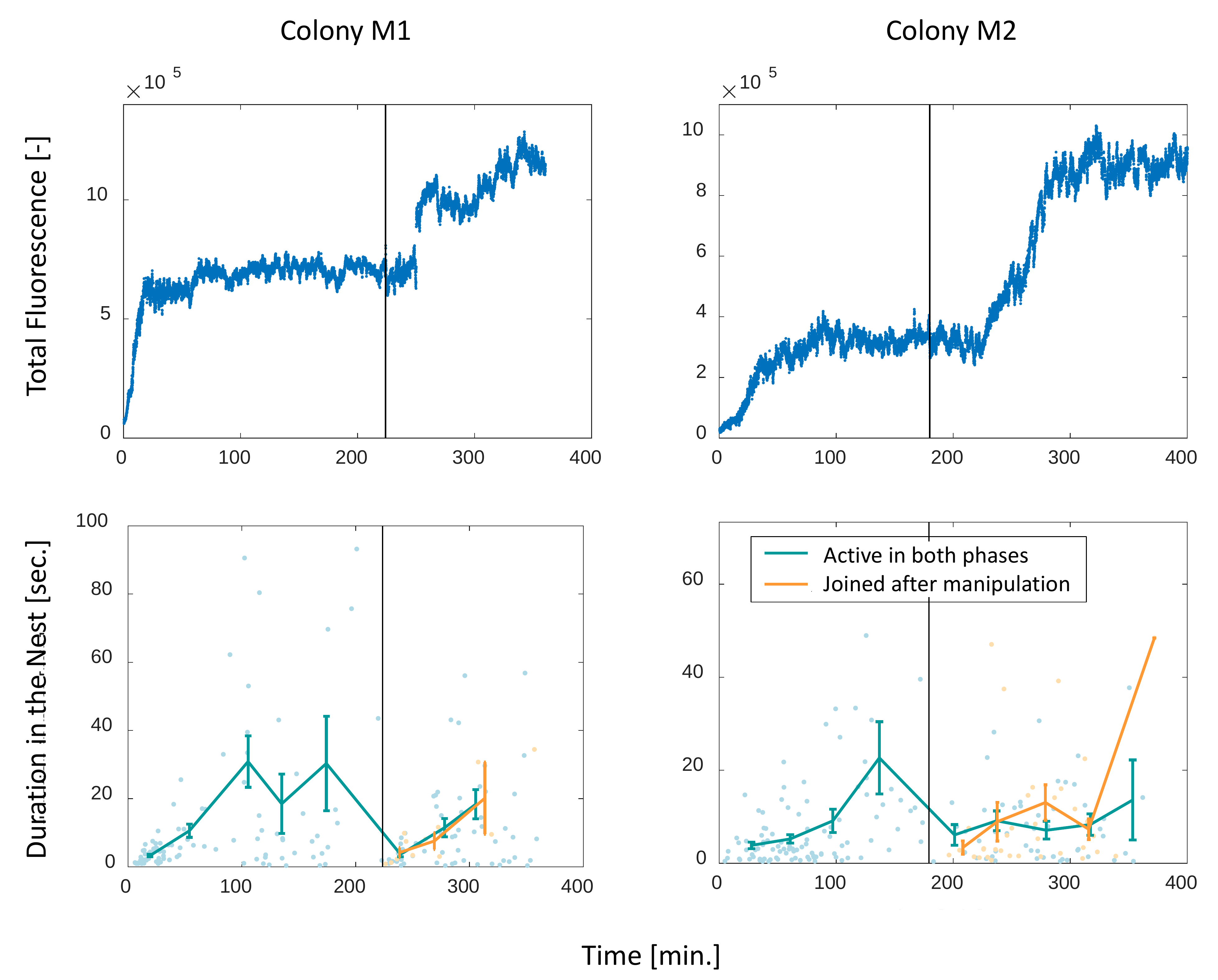}
    \caption{\textbf{Figure \ref{fig:cycles} - figure supplement 2: Perturbation Experiments.} Food accumulation, represented by the total fluorescence (top row), and durations of foragers in the nest (bottom row), as a function of time in the two experimental colonies that underwent colony state manipulation (see Methods). Solid black line represents the time of introducing new hungry ants to the system. The results of colony M2 are presented in Figure \ref{fig:cycles}d-e in the main text, where a detailed explanation may be found in the caption. Note that the plots depicting durations in the nest here differ from that in Figure \ref{fig:cycles}e, in that here durations are plotted by two separate groups: those of foragers which were active during the whole experiment (M1: n=47,13,15,13,12,16,17,14 for bins 1-8, respectively; M2: n=28,36,21,14,5,15,18,10,4, for bins 1-9, respectively) and those of foragers that began foraging after the manipulation (M1: n=6,3,2, for bins 1-3, respectively; M2: n=4,13,9,9,1, for bins 1-5, respectively). In Figure \ref{fig:cycles}e, all were pooled together. The observation that foragers of both groups displayed similar patterns after the manipulation highlights the causality of the effect of colony state on forager behavior as opposed to time or the forager's history.}
    \label{fig:manipSI}
\end{figure}

\begin{figure}
    \centering
    \includegraphics[scale=0.5]{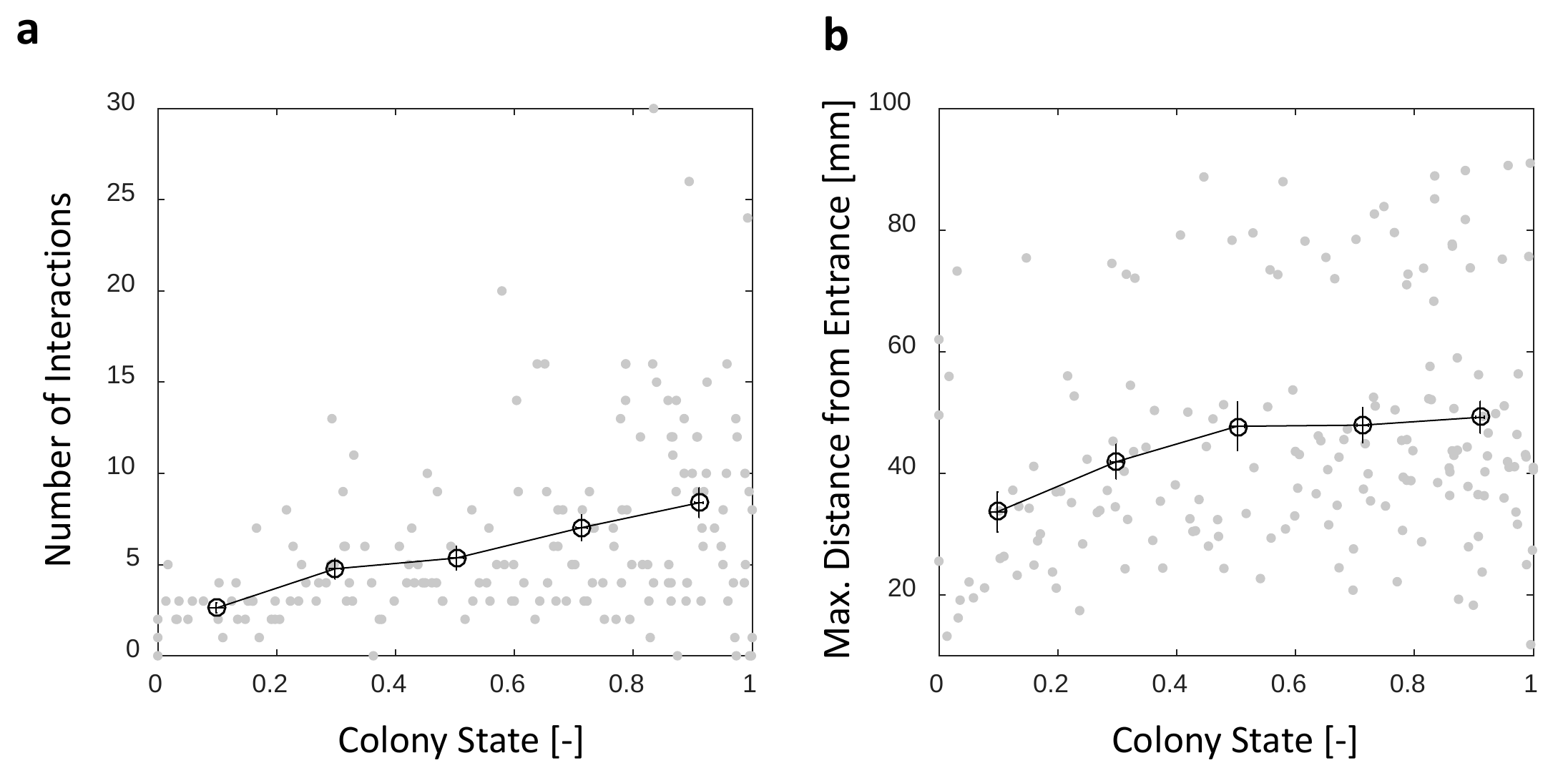}
    \caption{\textbf{Figure \ref{fig:cropatexit} - figure supplement 1: Number of Interactions and Maximal Distance from Entrance in a Forager's Visit in the Nest.} Both panels relate to the pooled data from all three observation experiments. \textbf{(a)} The number of interactions in which a forager participates in a single visit to the nest rises as the colony satiates. Raw data (gray) and mean $\pm$ SEM of binned data (black, n=26,26,28,39,57 for bins 1-5, respectively). \textbf{(b)} Foragers reach deeper locations in the nest as the colony satiates. Raw data (gray) and mean $\pm$ SEM of binned data (Black, n as in panel c). Data seems separated into clusters (upper and lower clouds) due to differences in the location of most of the ants within the nest between experiments.}
    \label{fig:exitcropSI}
\end{figure}

Our experimental method provided us with measurements of food volume in arbitrary units of fluorescent pixel intensity. Due to possible variations in lighting conditions between experiments, the obtained pixel intensities were incomparable. Therefore we used pixel intensities only for analyses perfomed within the same experiment (Fig. \ref{fig:global}b-c, Fig. \ref{fig:cycles}d-e and \ref{fig:globalSI}). For all other purposes, food load was estimated in normalized units. In analyses where individual crop loads and interaction volumes were linked to the global dynamics (Fig. \ref{fig:vdist}), absolute loads were important. Therefore, food loads were normalized between experiments, by dividing each measurement by the $90^{th}$ percentile crop load measurement of its experiment. In analyzing foragers' responses to their own crop loads (Fig. \ref{fig:P1P2P3}), the relative satiety state of each forager was of interest. Accordingly, food loads were normalized between foragers, by dividing the measurements of each forager by her own maximal measurement.

\subsubsection*{Exponential fits to interaction volume distributions}
Since exponential distributions could be fit only to `positive' interactions, i.e. where the forager was the donor, when we fit exponential distributions we neglected the negative interactions. Negative interactions constituted 216 out of 962, and accounted for 12\% of the total food flow. The consequence of this approximation is that we effectively lose 12\% accuracy in the modeled food flow. Despite this loss of accuracy, the results from this analysis were consistent with parameters obtained otherwise (without neglecting the negative interactions), ensuring that it was indeed sufficient to consider only the positive interactions.

\section*{Rich Media}

\subsubsection*{Video 1: A Trophallactic Event}
Two \textit{Camponotus sanctus} ants engaged in trophallaxis of fluorescent liquid food (presented in purple).

\subsubsection*{Video 2: Food Accumulation within a Colony of \textit{Camponotus sanctus}}
A starved colony replenishes on fluorescent liquid food (presented in red), brought in by few consistent foragers. Ant Identity is presented as a unique number next to her barcode tag.

\section*{Supplementary Figures}
\begin{itemize}
    \item Figure 1 - figure supplement 1: Food accumulation dynamics in all three experimental colonies
    \item Figure 1 - figure supplement 2: Food flows through individual foragers and their average
    \item Figure 2 - figure supplement 1: Interaction rate raw data
    \item Figure 2 - figure supplement 2: Interaction volume raw data
    \item Figure 2 - figure supplement 3: Unloading rate raw data
    \item Figure 3 - figure supplement 1: Interaction volume distributions given recipient crop load
    \item Figure 3 - figure supplement 2: Interaction volume distributions given forager crop load
    \item Figure 3 - figure supplement 3: Mean interaction volumes as a function of the recipient’s crop load.
    \item Figure 4 - figure supplement 1: Forager cycle times raw data
    \item Figure 4 - figure supplement 2: Perturbation experiments raw data
    \item Figure 5 - figure supplement 1: Number of interactions and maximal distance from entrance raw data
\end{itemize}

\section*{Source Data Files}
\begin{itemize}
    \item Figure 1 - source data 1: Trophallactic Interactions. This data also relates to Figures 2, 3 and 6.
    \item Figure 1 - source data 2: Temporal Data. This data also relates to Figures 2 and 6.
    \item Figure 2 - source data 1: Sensitivity analysis for binning interaction rate data.
    \item Figure 4 - source data 1: Foraging Cycles. This data also relates to Figure 5.
    \item Figure 4 - source data 2: Manipulation Experiments.
\end{itemize}

\section*{Acknowledgements}
We would like to acknowledge E. Segre, G. Han, for the technical help and Y. Dover and R. Harpaz for statistical advice. We also thank E. Fonio, Y. Heyman, A. Gelblum, H. Rajendran, A. Le Boeuf and T. Halperin for instructive comments on the manuscript.  O.F. is the incumbent of the Shloimo and Michla Tomarin Career Development Chair, was supported by the Israeli Science Foundation grant 833/15 and would like to thank the Clore Foundation for their ongoing generosity. This research is further supported by a research grant from the Estate of Rachmiel Ramon Bloch and by the European Research Council (ERC) under the European Unions Horizon 2020 research and innovation program (grant agreement No 648032).

\section*{Competing interests}
We confirm that this manuscript has not been published elsewhere and is not under consideration by any other journal. All of the authors agree with submission to eLife. We have no conflicts of interest to declare.

\bibliography{bibibi}

\beginsupplement

\pagebreak

\section*{Supplementary Information}

\end{document}